\documentclass{jpp}

\usepackage{epsfig}
\usepackage{enumerate}
\usepackage{graphicx}
\usepackage{natbib}
\usepackage{amssymb}
\def\Pm{\mbox{\rm P}_M}
\def\Rm{\mbox{\rm R}_M}
\def\Rmc{R_{\rm crit}}
\def\Rsat{R_{\rm sat}}
\def\Rey{\mbox{\rm Re}}
\def\apj{ApJ}

\def\mnras{MNRAS}
\def\aap{A\&A}
\def\apjl{ApJ}

\def\physrep{PhR}
\def\pre{PRE}

\def\nat{Natur}

\def\ssr{SSRv}

\newcommand{\be}{\begin{equation}}
\newcommand{\ee}{\end{equation}}

\title[Saturation of Zeldovich STF Map Dynamos]{Saturation of Zeldovich 
Stretch-Twist-Fold Map Dynamos}
\author[Amit Seta, Pallavi Bhat and Kandaswamy Subramanian]
{A\ls M\ls I\ls T\ns S\ls E\ls T\ls A\ls$^{1,2}$
  \thanks{Email address for correspondence: amitseta90@gmail.com},\ns
P\ls A\ls L\ls L\ls A\ls V\ls I\ns B\ls H\ls A\ls T$^3$\break
\and K\ls A\ls N\ls D\ls A\ls S\ls W\ls A\ls M\ls Y\ns S\ls U\ls B\ls R\ls A\ls M\ls A\ls N\ls I\ls A\ls N$^3$}

\affiliation{
$^1$UM-DAE Center For Excellence in Basic Sciences, University of Mumbai, Vidhyanagari Campus, Mumbai 400098, India\\
$^2$School of Mathematics \& Statistics, Newcastle University, 
Newcastle upon Tyne NE1 7RU, UK.\\[\affilskip]
$^3$IUCAA, Post Bag 4, Ganeshkhind, Pune 411 007, India.
}

\pubyear{}
\volume{}
\pagerange{}
\date{\today}

\begin{document}

\maketitle

\label{firstpage}

\begin{abstract}
Zeldovich's stretch-twist fold (STF) dynamo provided a breakthrough in conceptual
understanding of fast dynamos, including the small scale fluctuation dynamos. 
We study the evolution and saturation behaviour of two types
of generalized Baker's map dynamos, which have been used to model
Zeldovich's STF dynamo process.
Using such maps allows one to analyze dynamos at much higher magnetic Reynolds 
numbers $\Rm$ as compared to direct numerical simulations.
In the 2-strip map dynamo there is constant constructive folding
while the 4-strip map dynamo also allows the possibility of a destructive reversal of the 
field. Incorporating a diffusive step parameterised by 
$\Rm$ into the map, we find that the magnetic field 
$B(x)$ is amplified only above a critical $\Rm=\Rmc \sim 4$ for both 
types of dynamos. The growing $B(x)$ approaches 
a shape invariant eigenfunction independent of initial conditions, 
whose fine structure increases with increasing $\Rm$. 
Its power spectrum $M(k)$ displays sharp peaks
reflecting the fractal nature of $B(x)$ above the diffusive scale.
We explore the saturation of these dynamos in three ways; 
via a renormalized reduced effective $\Rm$ (Case I) 
or due to a decrease in the efficiency of the field amplification by 
stretching, without changing the map (Case IIa),
or changing the map (Case IIb), 
and a combination of both effects (Case III).
For Case I, we show that $B(x)$ in the saturated state, 
for both types of maps, approaches the marginal eigenfunction, 
which is obtained for $\Rm=\Rmc$ independent of the initial $\Rm=R_{M0}$. 
On the other hand in Case II, for the
2-strip map, we show that $B(x)$ saturates preserving
the structure of the kinematic eigenfunction. 
Thus the energy is transferred to larger scales in Case I
but remains at the smallest resistive scales in Case II as
can be seen from both $B(x)$ and $M(k)$. For the 4-strip map, 
$B(x)$ oscillates with time, although with
a structure similar to the kinematic eigenfunction.
Interestingly, the saturated state in Case III 
shows an intermediate behaviour,
with $B(x)$ similar to the kinematic eigenfunction at an intermediate 
$\Rm=\Rsat$, with $R_{M0} > \Rsat > \Rmc$. $\Rsat$ is determined by the 
relative importance of the increased diffusion versus the reduced stretching.
These saturation properties are akin to the range of possibilities
that have been discussed in the context of fluctuation dynamos.
\end{abstract}


\section{Introduction}
Magnetic fields in astrophysical systems are thought to arise by amplification of a 
seed magnetic field by dynamo action.
In this process kinetic energy of motions is converted to 
magnetic energy. A generic dynamo is the small-scale or fluctuation dynamo, which
arises in any random or turbulent flow 
\citep{Kaz68,ZRS83,Z90,HBD04,Schek04,BS05,TCB11,BSS12,BS14}. 
It is well known that when magnetic Reynolds number $\Rm$ of such a flow
is above a certain critical threshold $\Rmc \sim 100$, 
the magnetic field in the fluid is amplified rapidly on the
eddy turn over time-scales. This amplification 
is due to the random stretching by the velocity shear. Such shearing motions also 
lead to the magnetic field developing 
smaller and smaller spatial scales, until resistive diffusion becomes important 
to balance the growth. The field then becomes highly intermittent \citep{Z90}
with the kinematic eigenfunction having power peaked on the resistive scales
\citep{Kaz68}. 
For a random flow driven on a (single) scale $l$, the resistive
scale is $l_\eta \sim l/\Rm^{1/2}$, and for $\Rm \gg 1$, it is much
smaller than the driving scale $l$.
Eventually the Lorentz force of the growing magnetic field
provides a back reaction to the dynamo action,
leading to the saturation of magnetic field growth. The nature and spatial 
coherence of the field in the saturated state is of paramount importance
to the observational signatures of this field in 
different astrophysical systems \citep{SSH06,EV06,SC06,BS13},
but is however not well understood at present \citep{BS05,TCB11,BSS12}.

Indeed there is considerable evidence for
magnetic fields in several systems like 
galaxy clusters \citep{clarke_etal_01,Clarke04} and in young
galaxies \citep{Bernet08} from observations of Faraday rotation
that these systems induce on background polarized sources.
A possible origin of these fields is the fluctuation dynamo action.
However, whether one can indeed reproduce the observed levels
of Faraday rotation measure (FRM) depends on the spatial coherence
of the fields produced by the fluctuation dynamo.
As these systems have typically $\Rm \gg1$, the field
needs to become much more coherent in the saturated state 
than it is at the kinematic stage
for it to explain the observations \citep{BS13}.

The saturation of fluctuation dynamos has been studied via both direct
numerical simulations (DNS) and some simple analytical 
models. 
A simple model of \cite{S99} suggests that the dynamo can saturate
by the Lorentz force driving the dynamo to its marginal state. In 
such a case the magnetic field in the saturated state concentrates
on scales $l_c \sim l/\Rmc^{1/2}$. As $\Rmc \ll \Rm$ typically, this implies
a much more coherent field in the saturated state of the dynamo
than during the kinematic stage.
Using DNS with large magnetic Prandtl numbers 
($\Pm=\Rm/\Rey\gg 1$), but small fluid Reynolds numbers ($\Rey$),
\cite{Schek04} argued that the fluctuation dynamo saturates with the magnetic field  
still concentrated on resistive scales.
On other hand simulations of \cite{HBD03,HBD04} 
and \citet{Eyink13} with $\Pm=1$ and a large 
$\Rm=\Rey \approx 10^3$,
found that the magnetic integral scale is just a modest fraction of the 
velocity integral scale, and much larger than the resistive scale.
One could then expect significant FRMs, as is also consistent with 
the theoretical expectation from \citet{S99} and the DNS 
results of \citet{SSH06, CR09, BS13}.
The case when both $\Rey$ and $\Pm$ are large, as in 
galactic and cluster plasmas, is of course not easy to simulate
and the saturation of the fluctuation dynamo could be quite
different \citep{Eyink11}.

Note that DNS are limited by the $\Rm$ that are achievable
and still perhaps do not have a large enough $\Rm$ to unambiguously
determine the saturated state. At the same time the analytical
models are still rather simplistic. In this context one may
wonder if there is any other independent and simple way of studying
the generic saturation properties of fluctuation
dynamos. We consider here map dynamos that have been
studied earlier to mimic 
kinematic fast dynamo action, and examine how such maps could saturate.
Such map dynamos typically lead to a fractal structure
of the field, where the field goes into smaller and smaller scales \citep{FO88,FO90,STF95}.
However with the incorporation of a diffusive step
in the map, they can lead to eigenfunctions which preserve
their shape, and have the smallest scale determined
by the resistivity (or the effective $\Rm$). 
In the case of such map dynamos one can reach much larger
$\Rm$ than for the case of DNS. We then examine simple
models of saturating the map dynamo and study 
how the spatial structure of the map eigenfunction changes
from the kinematic to the saturated state. Our aim is then
to get insight into generic properties of
the saturated states of the fluctuation dynamo itself.  
 
In the next section we begin with the description of the
standard Stretch-Twist-Fold (STF) dynamo \citep{VZ72}. The corresponding map model
for the STF dynamo is outlined in section 3. Results from numerical
simulation of the STF map dynamo for various $\Rm$ are given in
section 4. The saturation of the STF map dynamos is taken up in section 5. 
The last section presents a discussion of the results and our conclusions.

\section{Zeldovich's STF dynamo}

To explain the possibility of the fast dynamos i.e. growth of magnetic field even when 
resistivity tends to zero, \cite{VZ72} put forward a heuristic description referred to 
as `Stretch-Twist-Fold' (STF) dynamo. The algorithm involves first stretching a closed flux 
tube to double its length 
(see for example Fig.~4.6 in \citet{BS05}) 
preserving its volume (a characteristic of an incompressible flow). 
Assuming magnetic flux to be frozen in the fluid, the magnetic field doubles 
as the area of the cross-section goes down by a factor of two.
Next, the flux tube is twisted into a figure eight 
and then folded so the direction of magnetic field is same in both sub-parts.
Then both parts are merged together into one 
through small diffusive effects to occupy the same volume as the starting flux tube. 
A weak diffusion is thus required to make the process of merging 
irreversible by smoothing region between the two flux tubes during 
joining without much loss in the flux or energy. 
\footnote{Note that diffusion is not essential for the process of amplification of 
magnetic field in the STF dynamo. 
Even in random flows where both constructive and destructive 
twisting and folding are possible, the probability of 
constructive effect dominates and leads 
to field amplification \citep{Z90,MRS84}. 
However without the diffusion, the map is in principle 
reversible, so that one can restore the previous state. 
This is not possible after the diffusive step. 
Of course diffusion is also required to develop an eigenfunction of the STF dynamo.}

It may thus be more appropriate to refer to this process as the Stretch-Twist-Fold-Merge dynamo,
although we continue with the standard terminology here.

Hence, the final flux tube becomes 
equivalent to the original single 
flux tube, but with a field that is double the initial field.
It is important to mark the way the two parts are folded: if they are folded 
with fields pointing in opposite directions
it would lead to the cancellation of the field and can model the field reversals 
due to turbulence in the actual physical situation.
For a constructive folding, with each step the flux and thus the magnetic field 
increases by a factor of $2$, repeating the same process $n$ times the magnetic 
field increases by a factor of $2^n$. Thus the growth rate is $\sim T^{-1} {\rm ln}2$ 
where $T$ is the time for one STF step. The stretching can also be done in a non-uniform manner. 
Suppose the stretching is done non-uniformly, by stretching 
a fraction $\beta$ (where $0<\beta<1/2$) of the circumference 
($2 \pi R$) by an amount 
$1/\beta$ and the remaining length $2 \pi \alpha R$ of the circumference by $1/\alpha$ (where $\alpha=1-\beta$).
Then this would give rise to the same growth rate in magnetic field as 
before but in a non-uniform way.
Repeated operation of the inhomogeneous STF process on a flux 
tube or on one that has developed reversals,
would lead to the magnetic field developing a fine scale structure 
that can mimic the intermittent nature of the field
generated by fluctuating dynamos. 

\section{Map Models for Zeldovich's STF Dynamos} 

\begin{figure}
\begin{center}
\includegraphics[width=12cm,height=6cm]{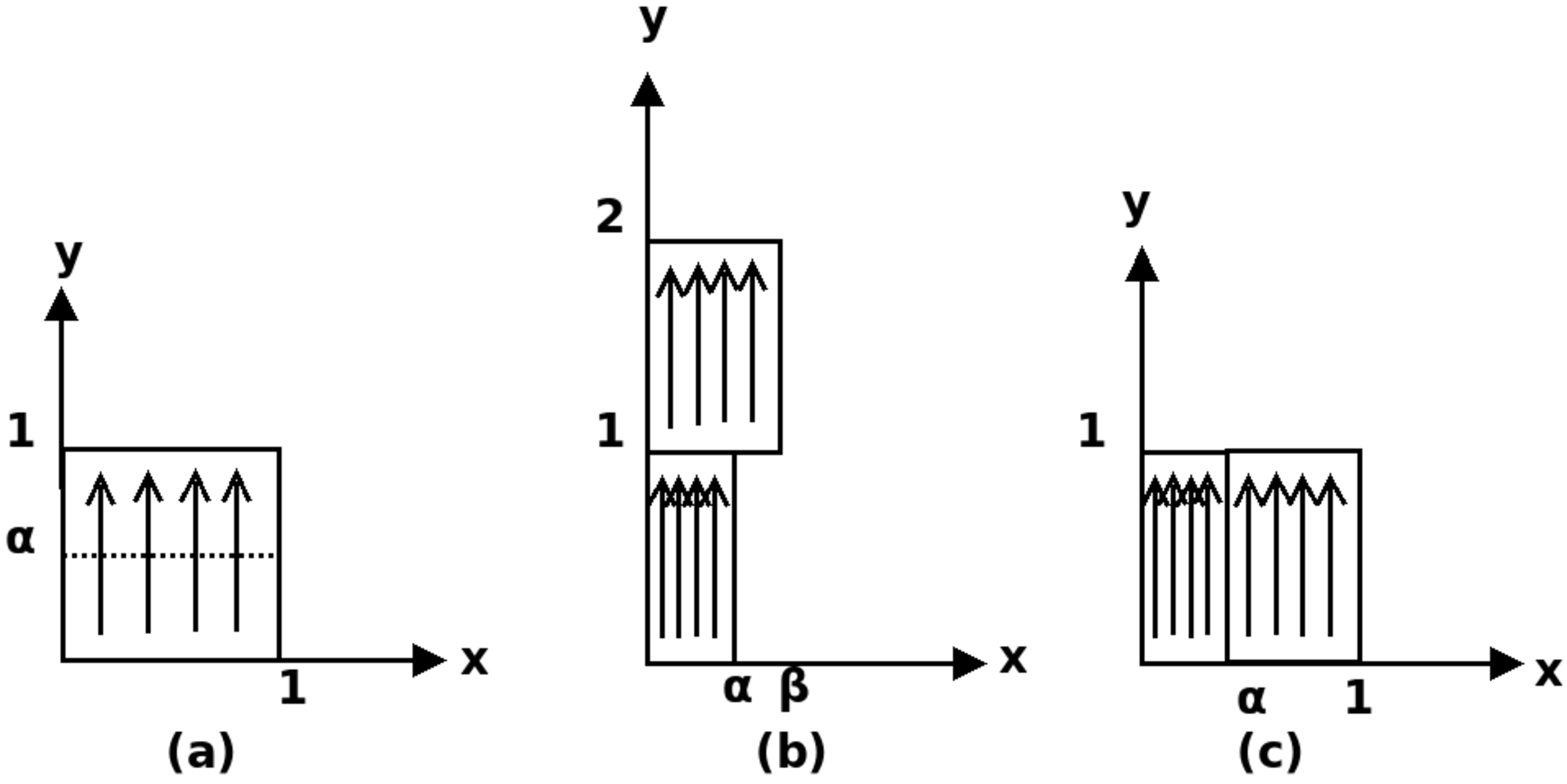} 
\caption{The two-dimensional Baker's map 
as a model for the 
stretch-twist-fold dynamo,  
but with non-uniform stretching.}
\end{center}
\label{fig.1}
\end{figure}

\cite{FO88, FO90} studied a map analogue of Zeldovich's Stretch-Twist-Fold 
fast dynamo. Fig.\ref{fig.1} represents the two-dimensional map which is used to model
the dynamo. This map is an example of the generalised Baker's map \citep{STF95}.
We begin with a perfectly conducting two-dimensional square sheet 
in the $(x,y)$ plane, and a uniform upward (or $y$) directed seed magnetic field
of say a unit strength.
The magnetic field initially, 
and at all times
is independent of $y$ (analogous to being independent of the toroidal direction 
in a flux tube). Now, the lower part of the square $(0 < y <\alpha)$ is horizontally 
compressed by a factor $\alpha$ and, to conserve the area simultaneously stretched 
vertically by factor $1/\alpha$. Similarly, the upper part $(\alpha < y <1)$ is compressed 
by factor $\beta$ (along $x$) and stretched by factor $1/\beta$ (along $y$) where 
$\beta = 1 -\alpha$. Then the two parts are separated, the magnetic field
between the two parts is cut and the two pieces are re-arranged to get back the original square. 
(This non-physical action allows one to describe an inherently 
three-dimensional physical process by a two-dimensional map; there cannot be any dynamo action in two-dimensional flows \citep{ZRS83}). 
As the flux is frozen in the region ($\eta \rightarrow 0 $),
the field through the $\alpha$ strip, $B_\alpha$ increases to 
$B/\alpha$ and the field through the $\beta$ strip, $B_\beta$ by 
$B/\beta$, where $B=1$ is the initial field. Then the total flux is $B \alpha + B \beta =2$, 
and thus the flux through entire square doubles. The average magnetic field 
also doubles within the square as compared to the initial field.

The map that captures the above process is as follows:
\begin{equation}
  x_{n+1} = \left\{
     \begin{array}{ll}
      \alpha x_n &: y_n < \alpha; \\ 
      \beta x_n + \alpha &:  y_n > \alpha; \\
     \end{array} \\
   \right. 
\label{eq.1} 
\end{equation}
\begin{equation}
 y_{n+1} = \left\{
     \begin{array}{ll}
       y_n/\alpha &: y_n < \alpha; \\ 
       (y_n - \alpha)/\beta &:  y_n > \alpha; \\
     \end{array}
   \right. 
\label{eq.2} 
\end{equation}
Note that $x_n$ and $y_n$ take values in the interval $[0,1]$. 
The corresponding amplification of magnetic field in the region is given as: 
\begin{equation} 
B_{n+1}(x_{n+1}) = \left\{
     \begin{array}{ll}
      B_n(x_n)/\alpha &: x_n < \alpha; \\ 
      B_n(x_n)/\beta &:  x_n > \alpha; \\
     \end{array}
   \right. 
\label{eq.3} 
\end{equation}

\begin{figure}
\begin{center}
\includegraphics[width=12cm,height=9cm]{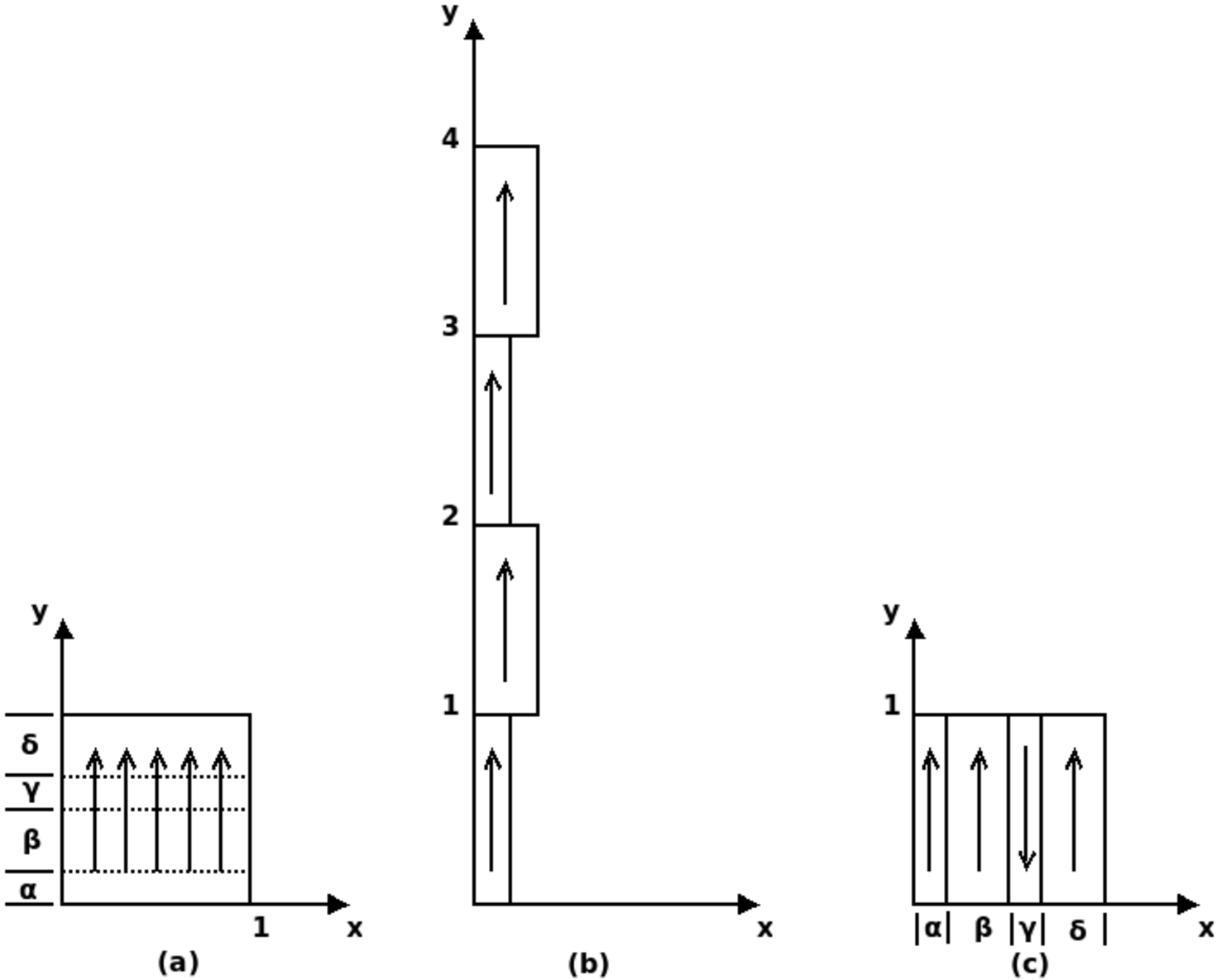} 
\caption{The two-dimensional 4-strip Baker's map 
as a model for the 
stretch-twist-fold dynamo 
with cancellation of fields.}
\label{4stripdesc}
\end{center}
\end{figure}
\subsection{Amplification, cancellation and the four-strip map}

To include field reversals, which are an important part of the 
physical dynamo process leading to cancellation of magnetic fields,
\cite{FO88,FO90} suggest a different model for STF dynamos. 
The flux tube in this case is stretched non-uniformly to four 
times its original circumference,
twisted in to four loops,
and two of the loops are folded
with the same orientation of magnetic field while the other two with 
the opposite directions of fields thus leading to a partial field cancellation. 
The corresponding map is shown in Fig.~\ref{4stripdesc}.
The net field increase is same as in the two-strip map, the field doubles 
with each step. As shown in the Fig.~\ref{4stripdesc}, the analogous map 
would involve dividing the square into four strips and
while rearranging the strips, one of them is inverted and 
then placed to regain the initial square configuration.
\begin{figure}
  \includegraphics[width=6.5cm,height=5.5cm]{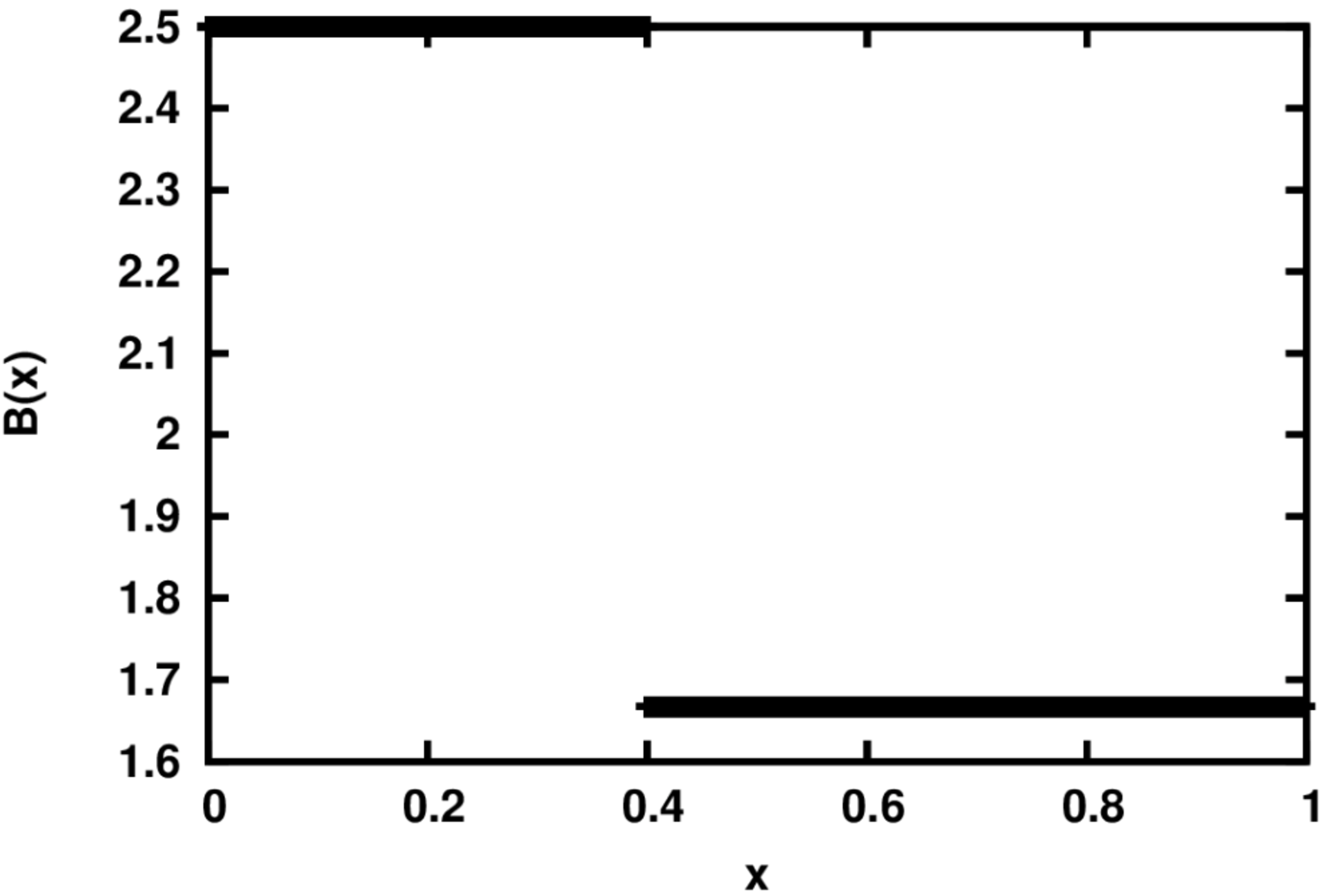}
    \includegraphics[width=6.5cm,height=5.5cm]{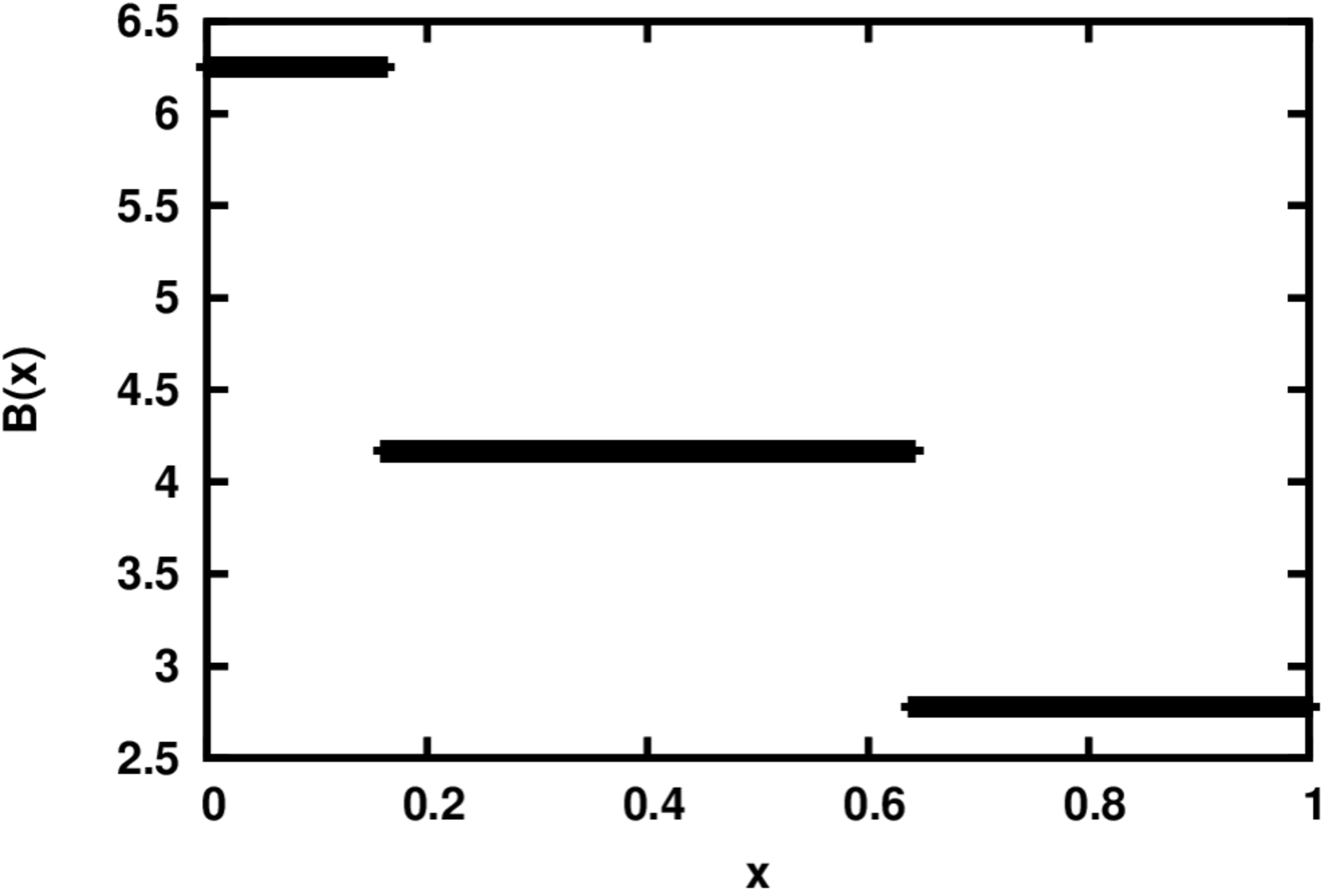}
\\
     \includegraphics[width=6.5cm,height=5.5cm]{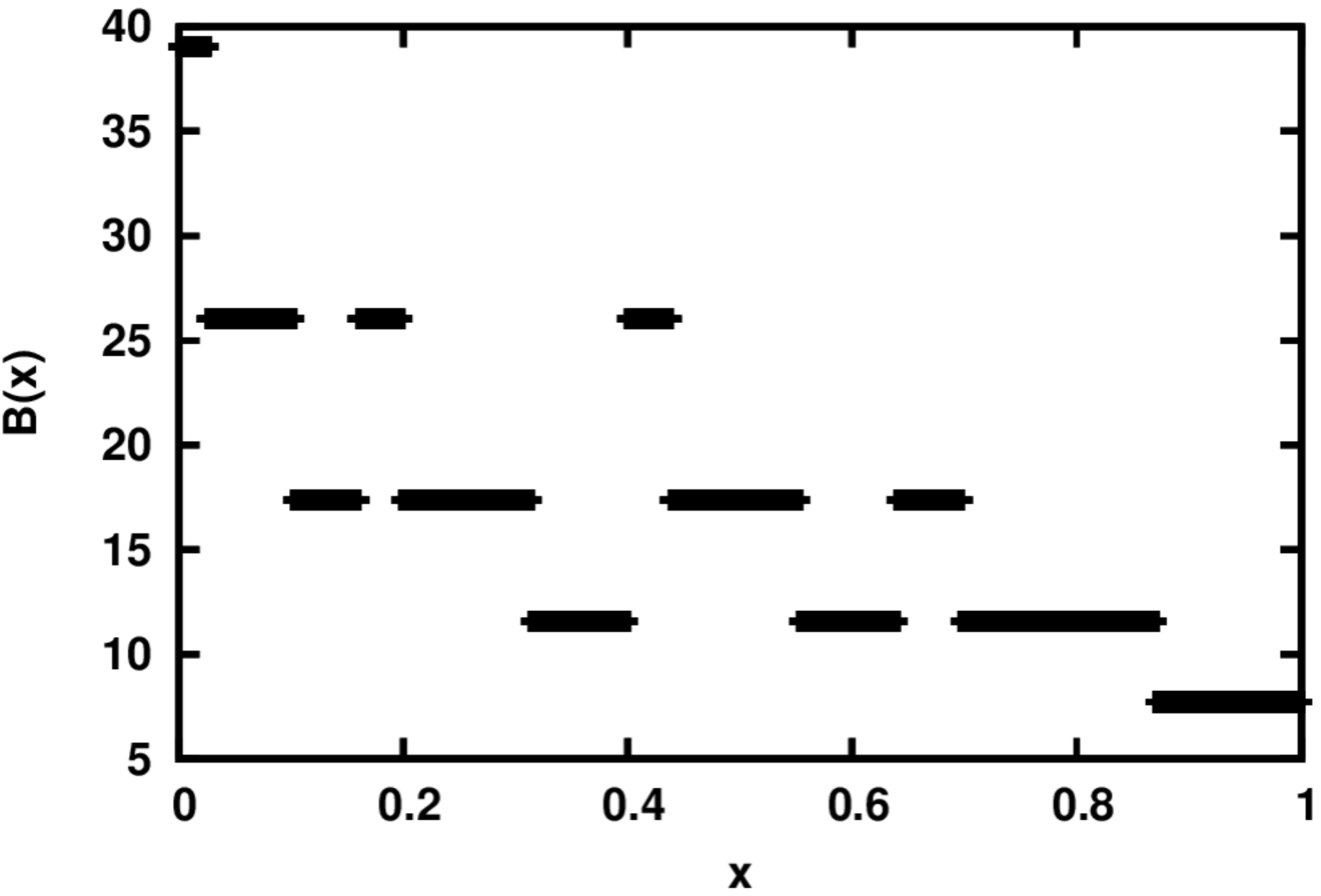}
       \includegraphics[width=6.5cm,height=5.5cm]{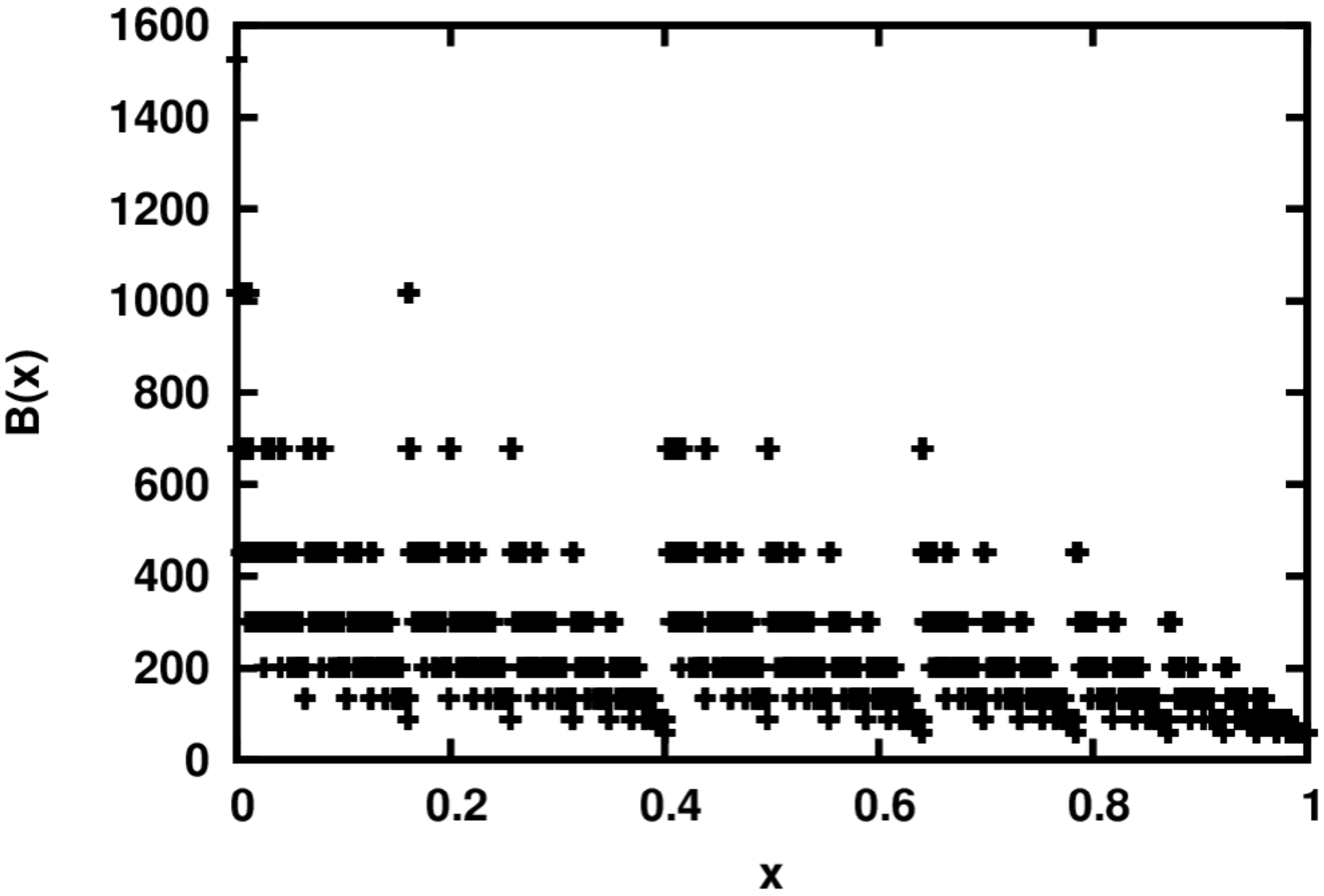}
 \caption{Magnetic field evolution in the two-strip map with $\alpha = 0.4$ 
after 1, 2, 4 and 8 iterations (top left to bottom right panels). 
The magnetic field strength $B(x)$ is given in the units of the seed field.
We see the amplification of the field strength and development 
of the intermittent structure
of the field as the number of iterations increases.}
\label{rmevol}
\end{figure}
The analytical description of the four-strip map is given by-
\begin{equation}
x_{n+1} = \left\{
     \begin{array}{ll}
      \alpha x_n &: y_n < \alpha; \\
      \beta x_n + \alpha &: \alpha < y_n < (\alpha+\beta); \\ 
      \gamma(1-x_n) + \alpha + \beta &: (\alpha+\beta) < y_n < (\alpha+\beta+\gamma); \\
      \delta x_n + \alpha+\beta+\gamma &: (\alpha+\beta+\gamma) < y_n < 1;  \\
     \end{array}
   \right.
\label{eq.4}
\end{equation}
\begin{equation}
   y_{n+1} = \left\{
     \begin{array}{ll}
       y_n/\alpha &: y_n < \alpha; \\
      (y_n - \alpha)/\beta &: \alpha < y_n < (\alpha+\beta); \\ 
      ((\alpha + \beta + \gamma)-y_n)/\gamma&: (\alpha+\beta) < y_n < (\alpha+\beta+\gamma); \\
      (y_n - (\alpha + \beta + \gamma))/\delta &: (\alpha+\beta+\gamma) < y_n < 1;  \\
     \end{array}
   \right.
\label{eq.5}
\end{equation}
Again $x_n$ and $y_n$ take values in the interval $[0,1]$.
The corresponding amplification of magnetic field with a flip of sign in the third strip is now given by:
\begin{equation} 
B_{n+1}(x_{n+1}) = \left\{
     \begin{array}{ll}
      B_n(x_n) /\alpha &: y_n < \alpha; \\
      B_n(x_n) /\beta &: \alpha < y_n < (\alpha+\beta); \\ 
      - B_n(x_n) /\gamma&: (\alpha+\beta) < y_n < (\alpha+\beta+\gamma); \\
      B_n(x_n)  / \delta &: (\alpha+\beta+\gamma) < y_n < 1;  \\
      \end{array}
   \right.  
\label{eq.6}
\end{equation}

We will study both two-strip and four-strip maps in what follows.

\begin{figure}
\begin{center}
 \includegraphics[width=6.5cm,height=5.5cm]{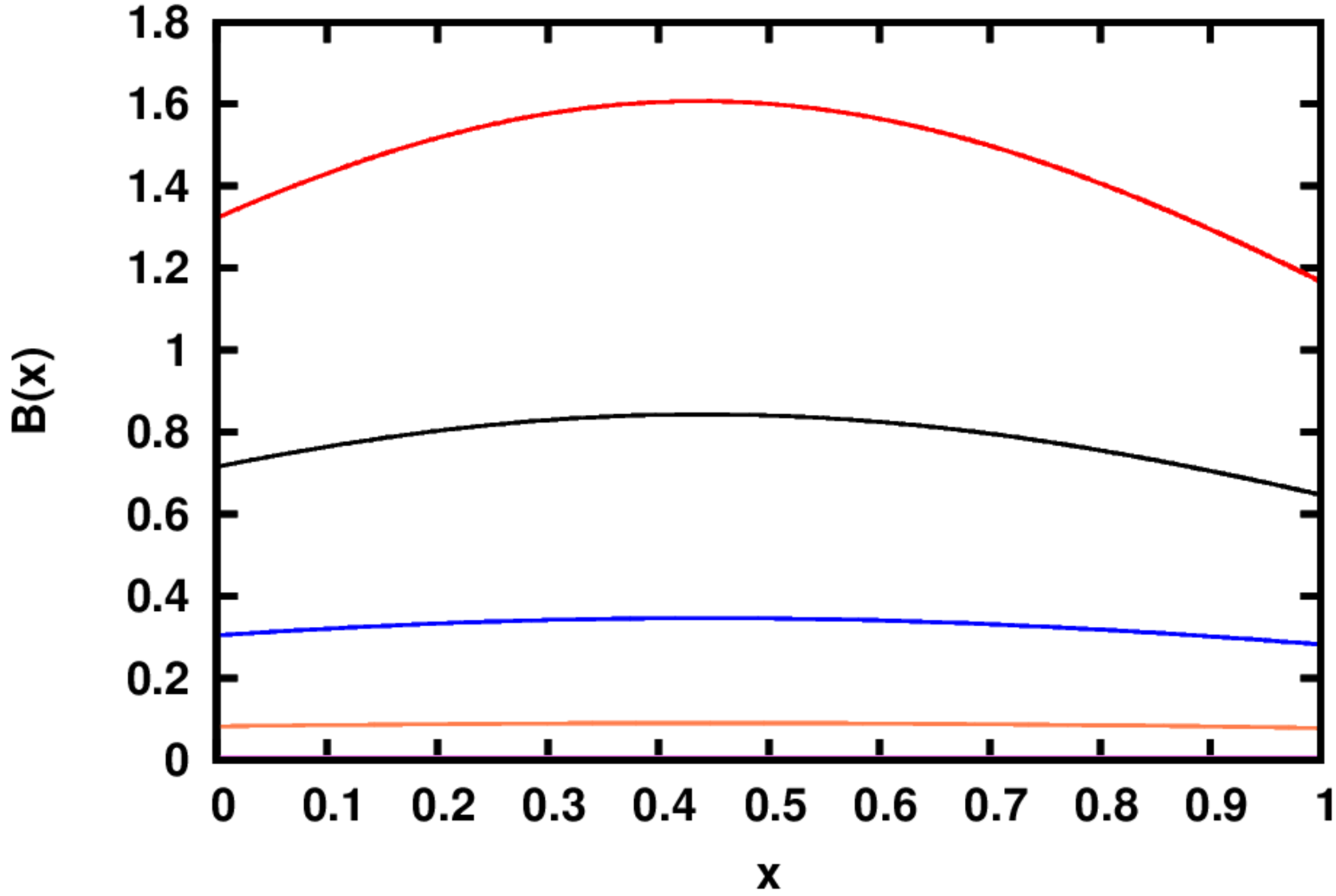}
 \includegraphics[width=6.5cm,height=5.5cm]{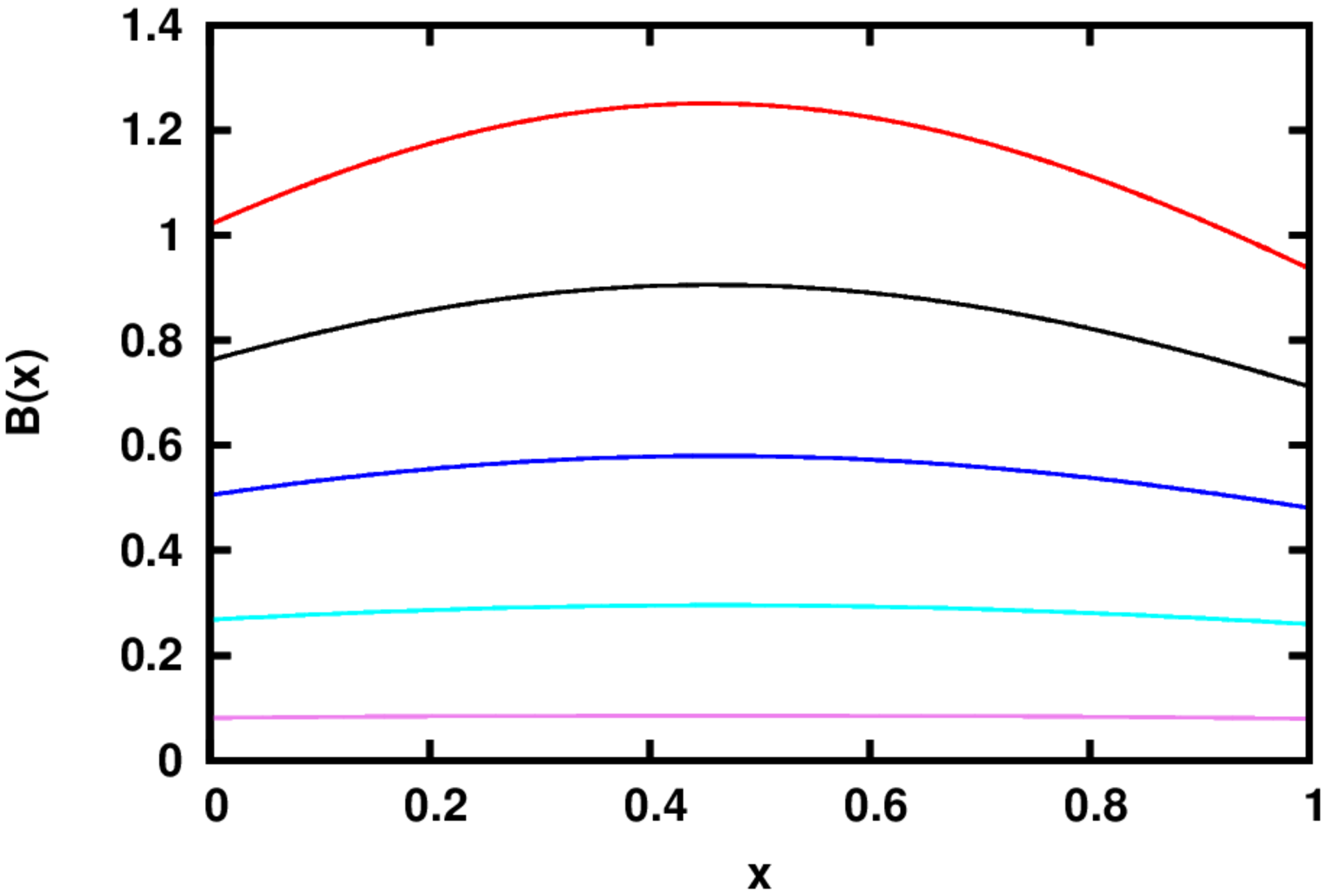}
 \caption{$B(x)$ for the two-strip map (left panel) and for the four-strip map 
(right panel) including diffusion, after 8 iterations. 
In the left panel the curves from bottom to top correspond to values of 
$\Rm = 2, 3, 4, 5$ respectively, while in the right panel they
correspond to $\Rm=1, 2, 3, 4, 5$. The field is amplified for a
critical magnetic Reynolds number $\Rmc \gtrsim 4$ in both cases.}
 \label{rmcrit}
\end{center}
\end{figure}
\begin{figure}
 \includegraphics[width=6.5cm,height=5.5cm]{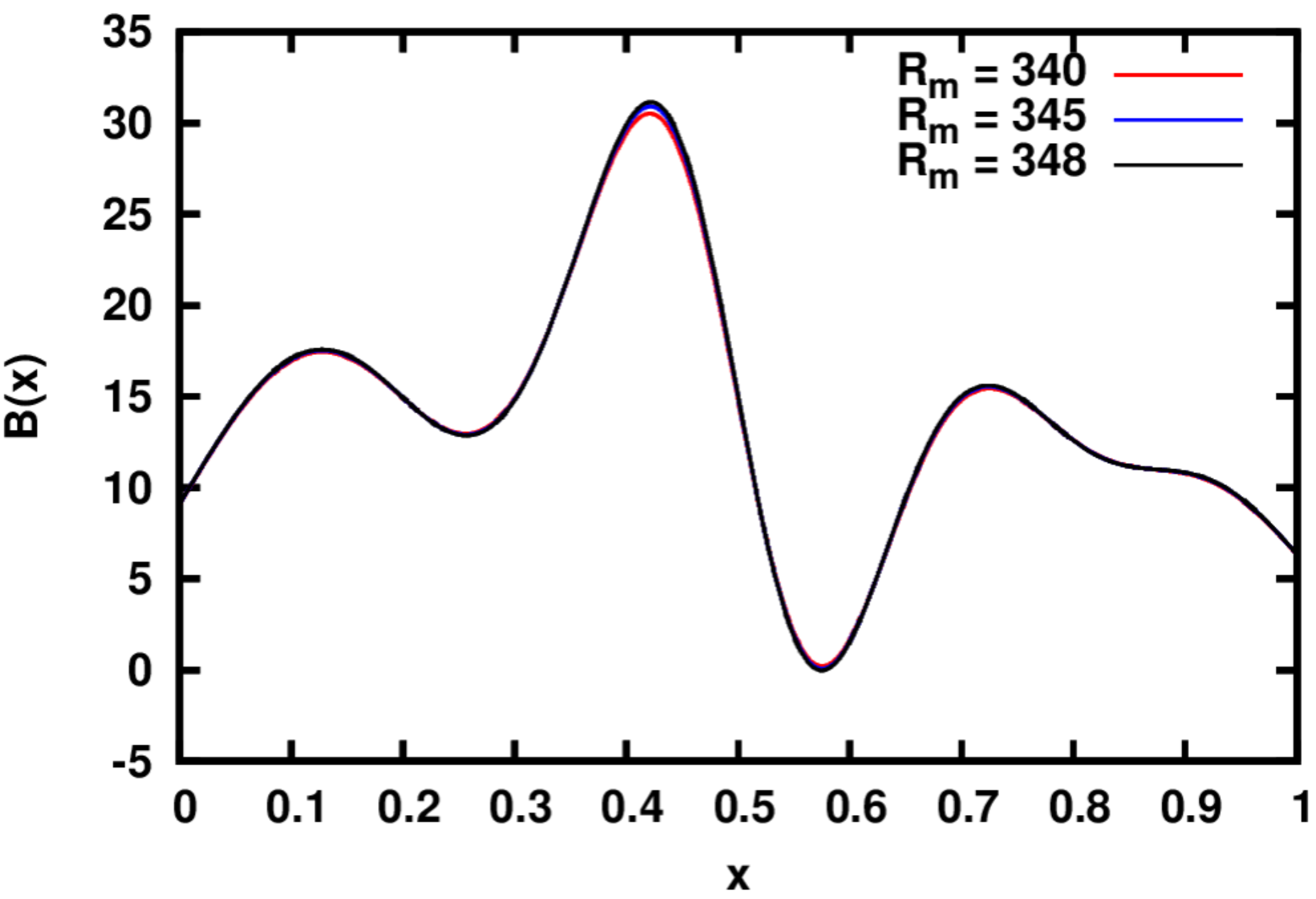}
 \includegraphics[width=6.5cm,height=5.5cm]{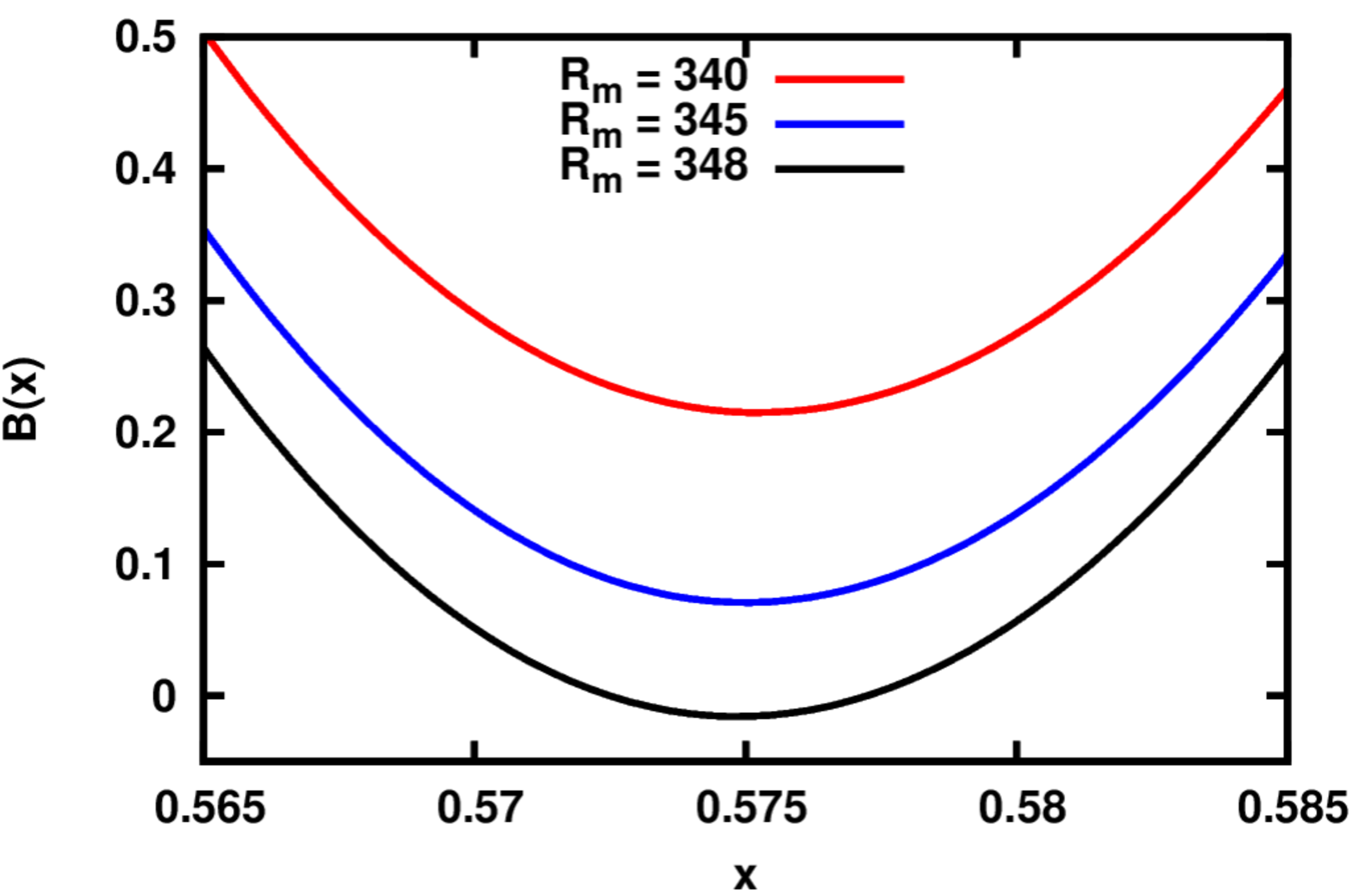}
 \caption{Left-hand panel: $B(x)$ for the four-strip map after 4 iterations, with the 
darkest curve corresponding to $\Rm = 348$. 
A magnified version is shown in the right panel. 
$B(x)$ becomes negative first for 
$\Rm \sim 348$, depending on the stretching parameters. Here we have adopted 
$\alpha = \delta = 7/16$ and $\beta = \gamma = 1/16$.}
 \label{4stripNeg}
\end{figure}
 
\subsection{Magnetic Diffusion}

As discussed above, 
it is important to include the smoothing effect of
diffusion in the STF dynamo.
We do this in the maps by convolving the evolved magnetic field
after each step with a Gaussian. 
If $T$ is the time interval for each complete cycle of the STF map, 
the assumption is that for the time $T/2$ the diffusive term in the induction
equation can be neglected, the flux is frozen and the field amplifies. 
For the remaining period, $T/2$, the advection term
vanishes and diffusion acts, 
with twice the normal diffusivity.
The magnetic Reynolds number $\Rm$ is the measure of 
advection in comparison to diffusion
and thus should be a parameter in the convolving function. 
The convolving Gaussian (also the Green's function for diffusion) is therefore taken as \citep{FO90},
\begin{equation}
 G(x,x') = (\Rm/ 4 \pi T)^{1/2} exp[-(x-x')^2 (\Rm/4T)]. 
\label{convolve}
\end{equation}
The width of the Gaussian ($\sigma = \sqrt{2T/\Rm}$) is inversely
proportional to $\sqrt{\Rm}$ and shows that the diffusion 
diminishes when $\Rm$ increases and viceversa.
A single iteration  involves applying the map to amplify the magnetic field and then 
convolving the evolved field with the above Gaussian.

\section{Results: Kinematic Stage}

We have coded the amplification of the magnetic field using maps given 
in Eq.~(\ref{eq.1})-(\ref{eq.6}) and also included diffusion using the set 
of points in the unit square in the $(x,y)$ plane. The positions of the points are evolved using the map equations. 
The distribution of points in the interval $[0,1]$ is dynamic and the map automatically allocates more points 
to areas where the magnetic field has finer structure. This allows us to 
achieve higher $\Rm$ than the number of points that are used. 
We have generally adopted $300 \times 300$ points for lower $\Rm$ 
runs while for higher $\Rm$ we use $500 \times 500$ points. 

We first show in Fig.~\ref{rmevol}, the result of applying the two strip map, 
excluding diffusion to explore kinematic evolution of magnetic field 
starting from a unit seed field. Fig.~\ref{rmevol} shows the magnetic field $B(x)$, as a function of 
position $x$, after 1, 2, 4 and 8 iterations. We have adopted $\alpha 
= 0.4$. It is clear from Fig.~\ref{rmevol} that as the number of iterations 
increases magnetic field grows but becomes more and more intermittent. 
Moreover, within a small spatial scale the magnetic field varies
significantly and develops a fractal structure \citep{FO88, FO90}.

\begin{figure}
\begin{center}
 \includegraphics[width=12cm,height=7cm]{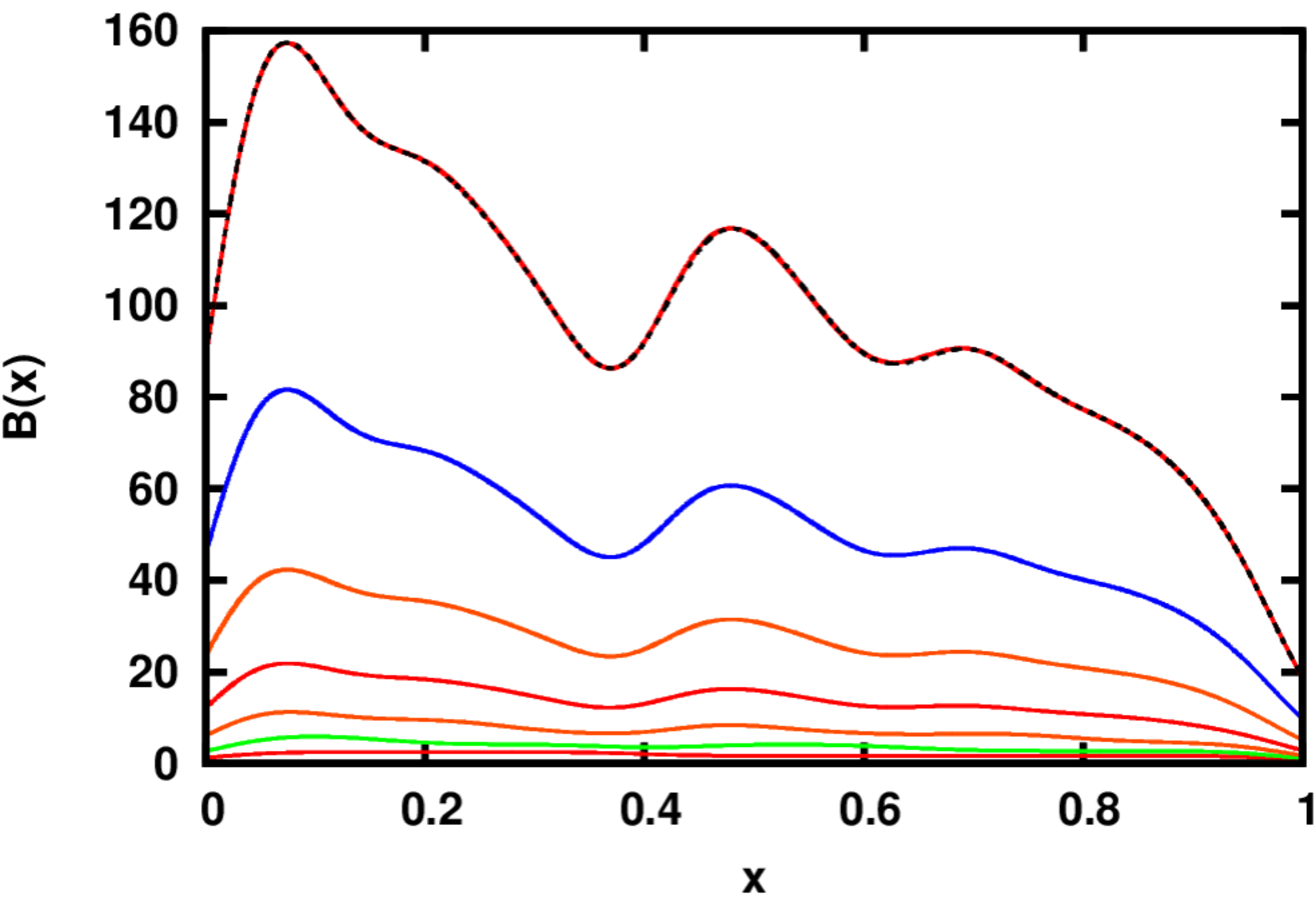}
 \caption{$B(x)$ for the 2-strip map for iterations 1 to 7 (from bottom to top),
adopting $\alpha = 0.4$ and $\Rm = 1000$. $B(x)$ after the 7th iteration 
is just a scaled version of $B(x)$ after the 6th one illustrating 
the development of the map eigenfunction.}
 \label{kinevol}
\end{center}
\end{figure}
\begin{figure}
  \includegraphics[width=6.5cm,height=5.5cm]{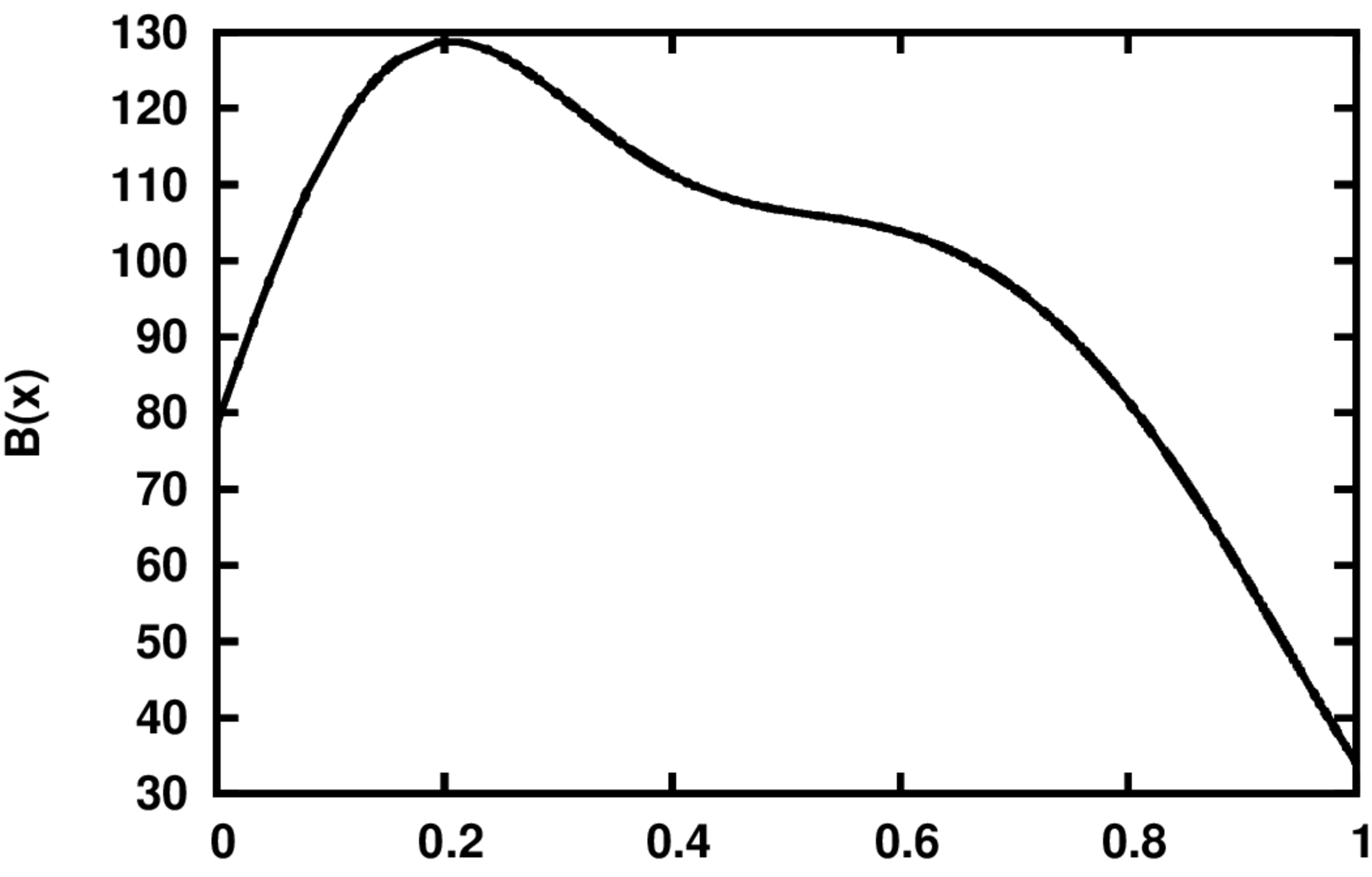}
    \includegraphics[width=6.5cm,height=5.5cm]{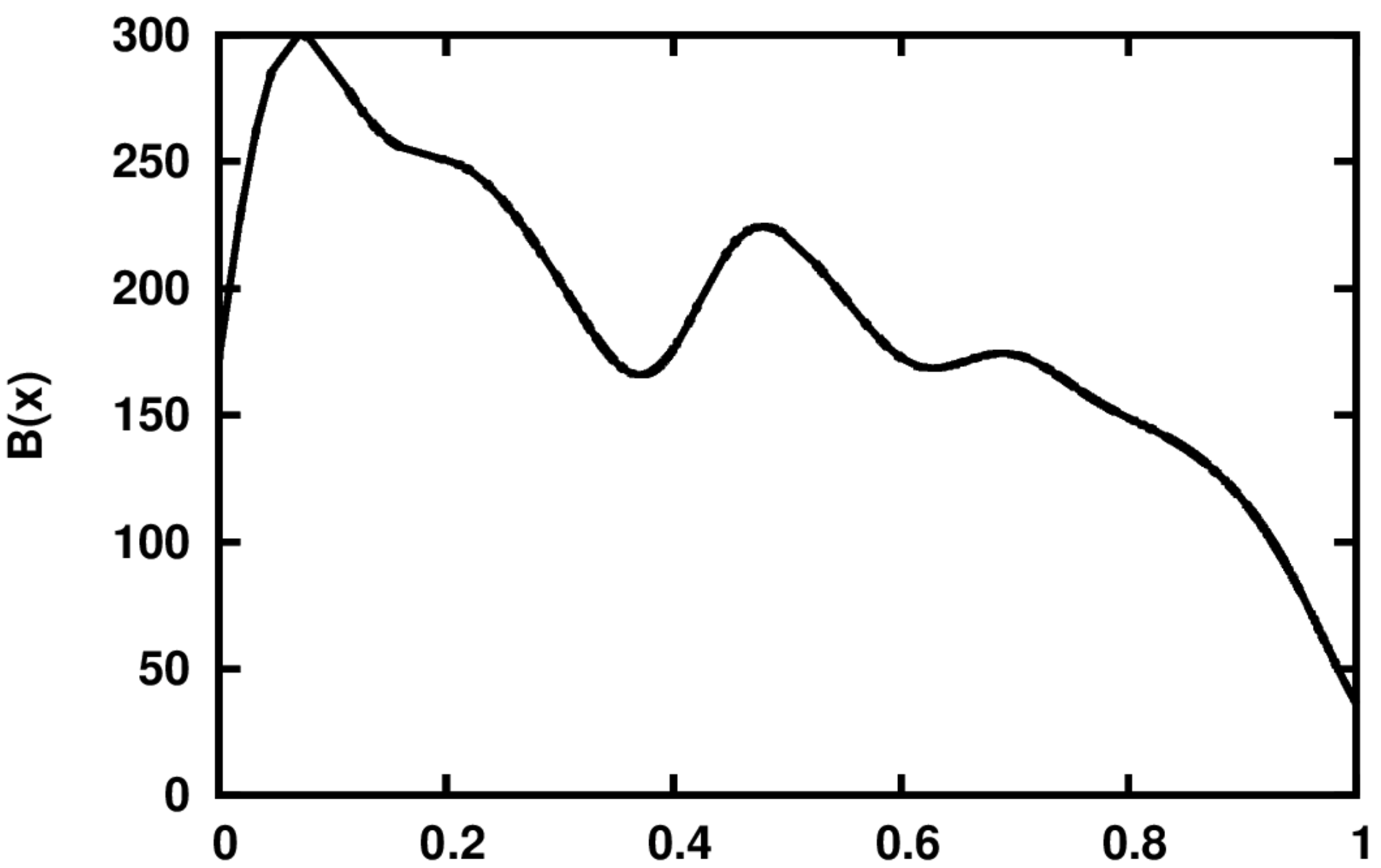}
\\ 
     \includegraphics[width=6.5cm,height=5.5cm]{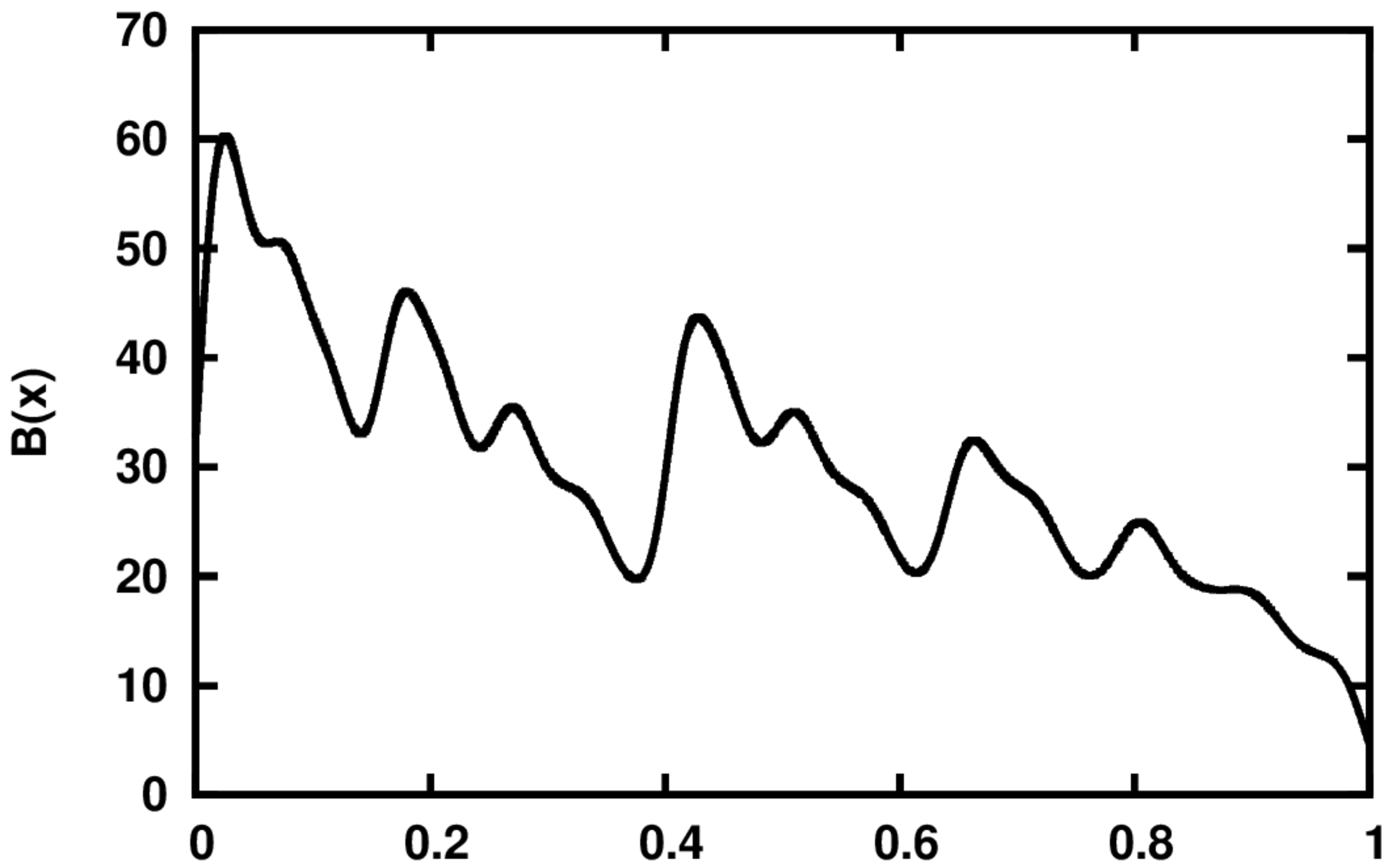}
       \includegraphics[width=6.5cm,height=5.5cm]{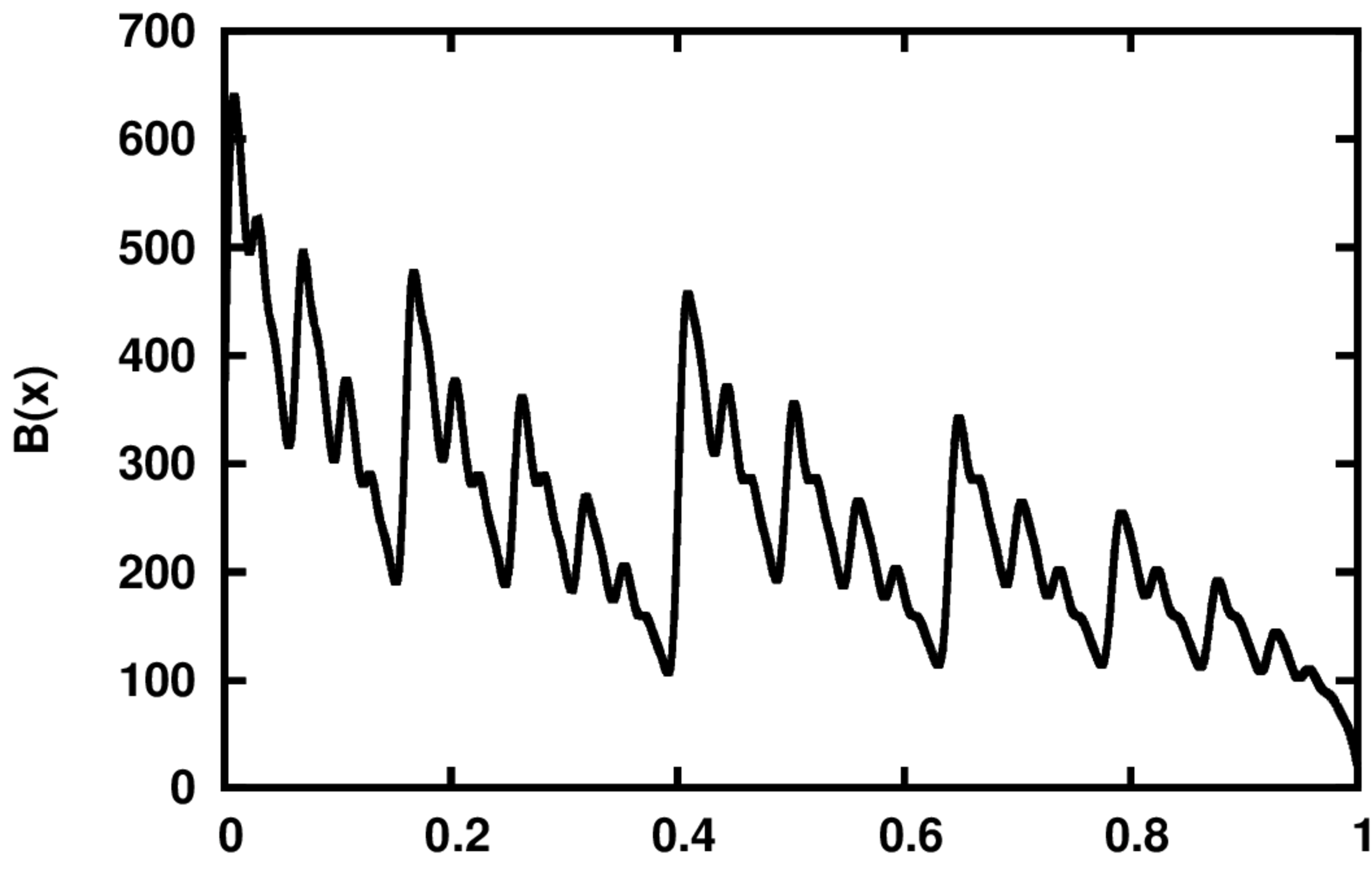}
 \caption{Eigenfunctions for the 2-strip map with $\alpha=0.4$ and 
for $\Rm = 100,1000,10^4,10^5$ (top left to bottom right panels).}
\label{2strip}
 \end{figure}
\subsection{Critical magnetic Reynolds number}
In the presence of diffusion, the amplification has to win over it for the net amplification to occur.
This introduces a critical $\Rm = \Rmc$, only above which the net 
amplification takes place.
We show in Fig.~\ref{rmcrit} (left panel) 
that starting with an initial seed field of 1, 
$B(x)$ is not amplified unless $\Rm > \Rmc$.
The critical magnetic Reynolds number for the two-strip map is $\Rmc \approx 4.35$.

We have also incorporated diffusion into the four-strip map with cancellation adopting $\alpha = \delta = 7/16$ and $\beta = \gamma = 1/16$.
The results which are shown in Fig.~\ref{4stripNeg} (right panel) 
suggest a critical magnetic Reynolds number which is very close to 
that obtained for the two-strip map.

It is interesting to note that, in the four-strip map, where the magnetic field
of both polarities are amplified, there is no field with negative polarity in the solution 
till a large enough $\Rm$ is reached. This critical $\Rm$ of course 
depends on the stretching parameters, in particular the value of $\gamma$, which 
determines the degree of amplification of the negative field. 
As shown in Fig.~\ref{4stripNeg}, the field in the region has negative polarity 
solutions only for $\Rm \gtrsim 348$ for $\alpha = \delta = 7/16$ and $\beta = \gamma = 1/16$.

\subsection{Eigenfunctions of the map dynamo}

After a few iterations, the function $B(x)$ representing variation of magnetic field along $x$ settles to an eigenfunction
of a specific shape and $B(x)$ from future iterations can be matched to it by scaling.
This can be seen in Fig.~\ref{kinevol} for the two-strip map. 
For example $B(x)$ after the $7$th iteration 
can be scaled back in amplitude to $B(x)$ after the $6$th iteration. 
Thus $B(x)$ tends to an 
eigenfunction of the map with diffusion included (which represents the STF dynamo). 
The eigenfunction of the two-strip map depends of course on $\Rm$.
\begin{figure}

  \includegraphics[width=6.5cm,height=5.5cm]{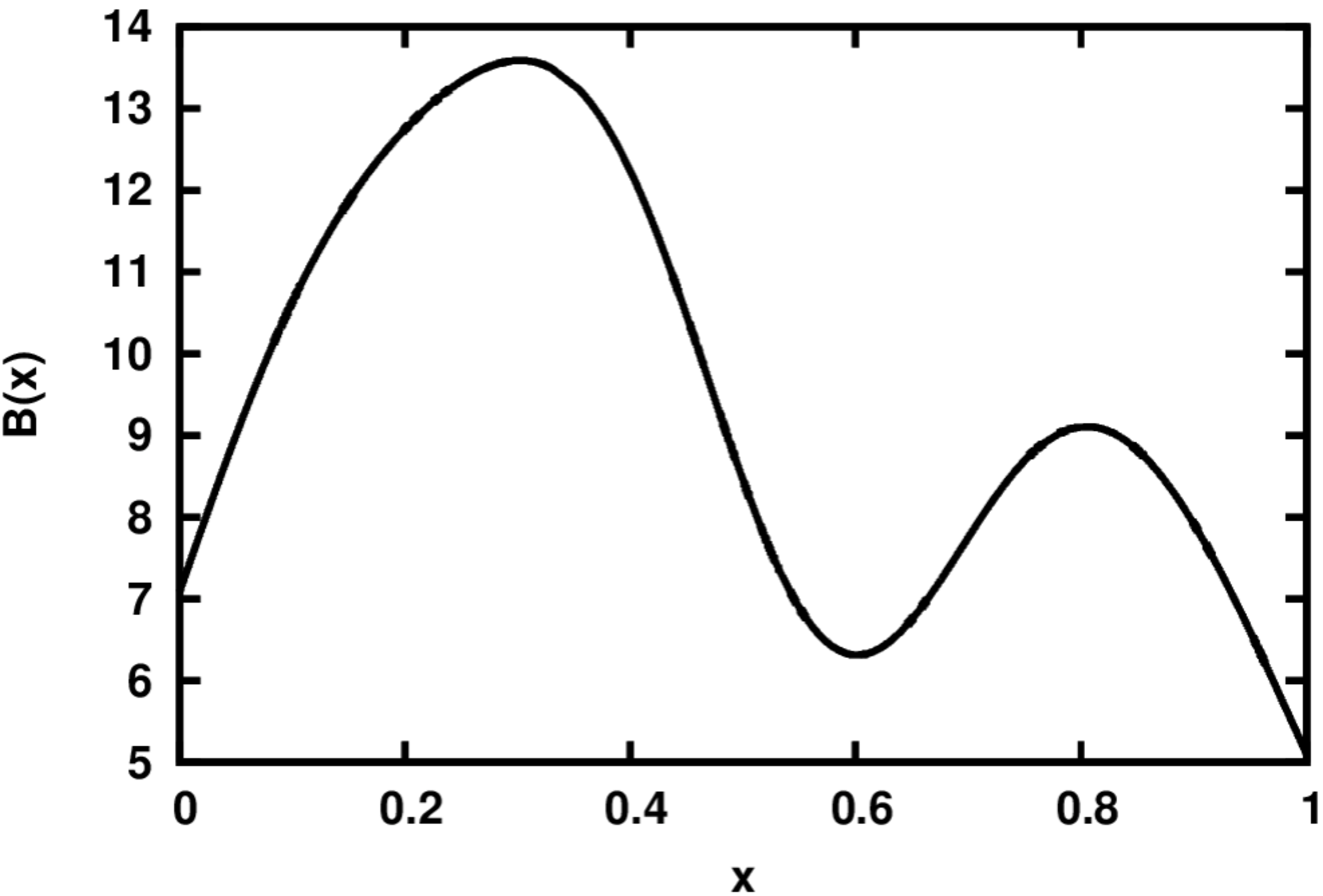}
    \includegraphics[width=6.5cm,height=5.5cm]{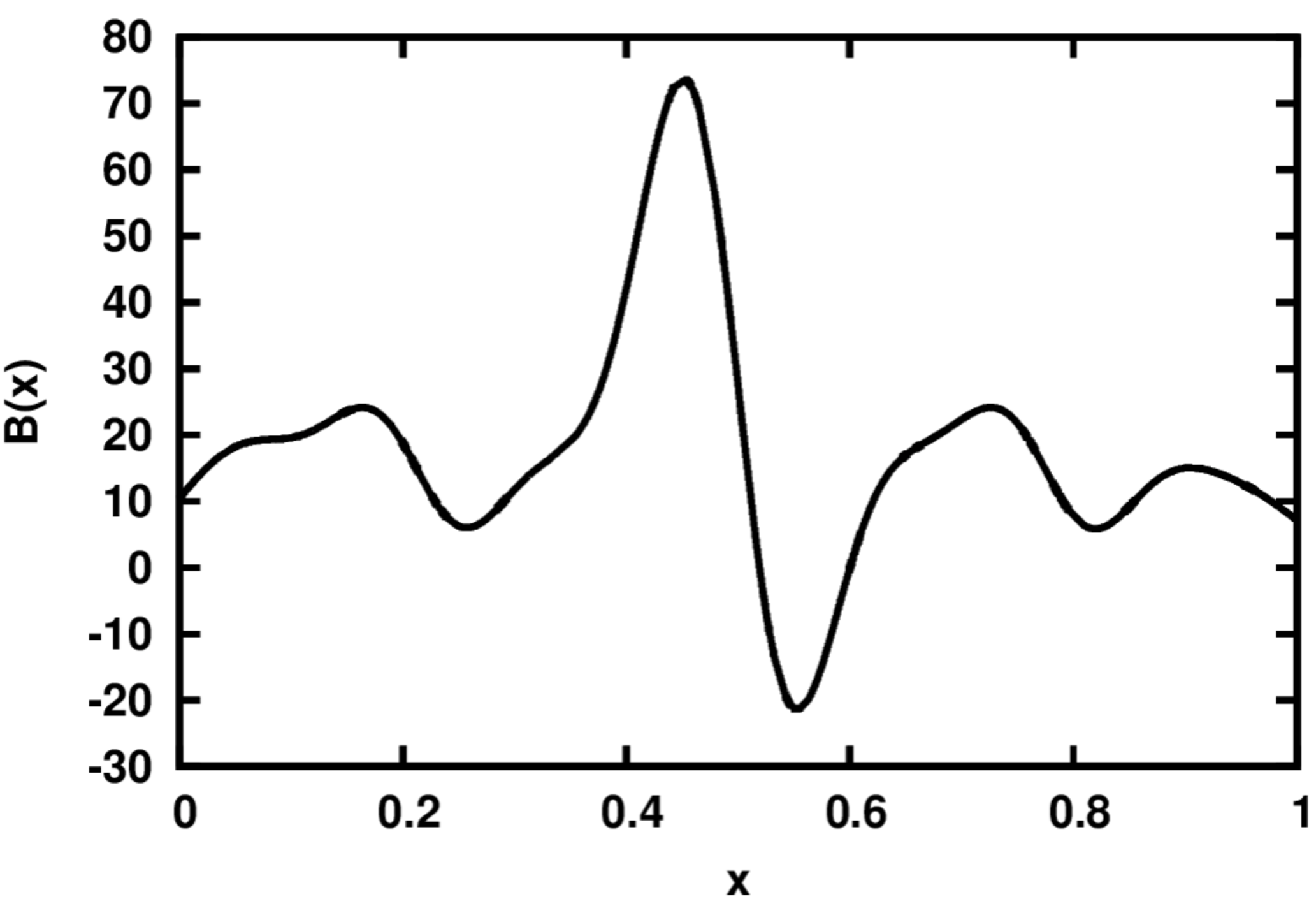}
\\
     \includegraphics[width=6.5cm,height=5.5cm]{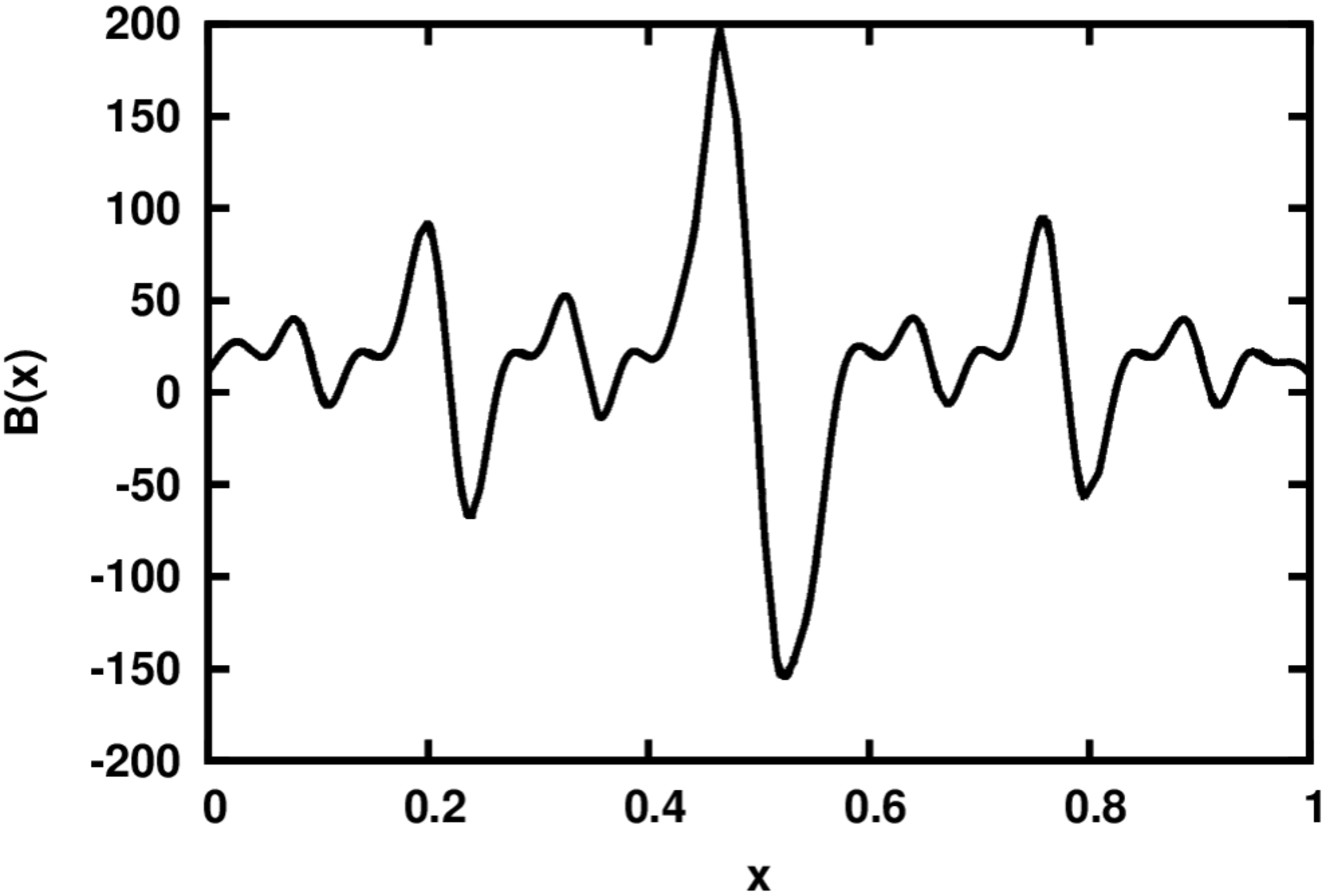}
       \includegraphics[width=6.5cm,height=5.5cm]{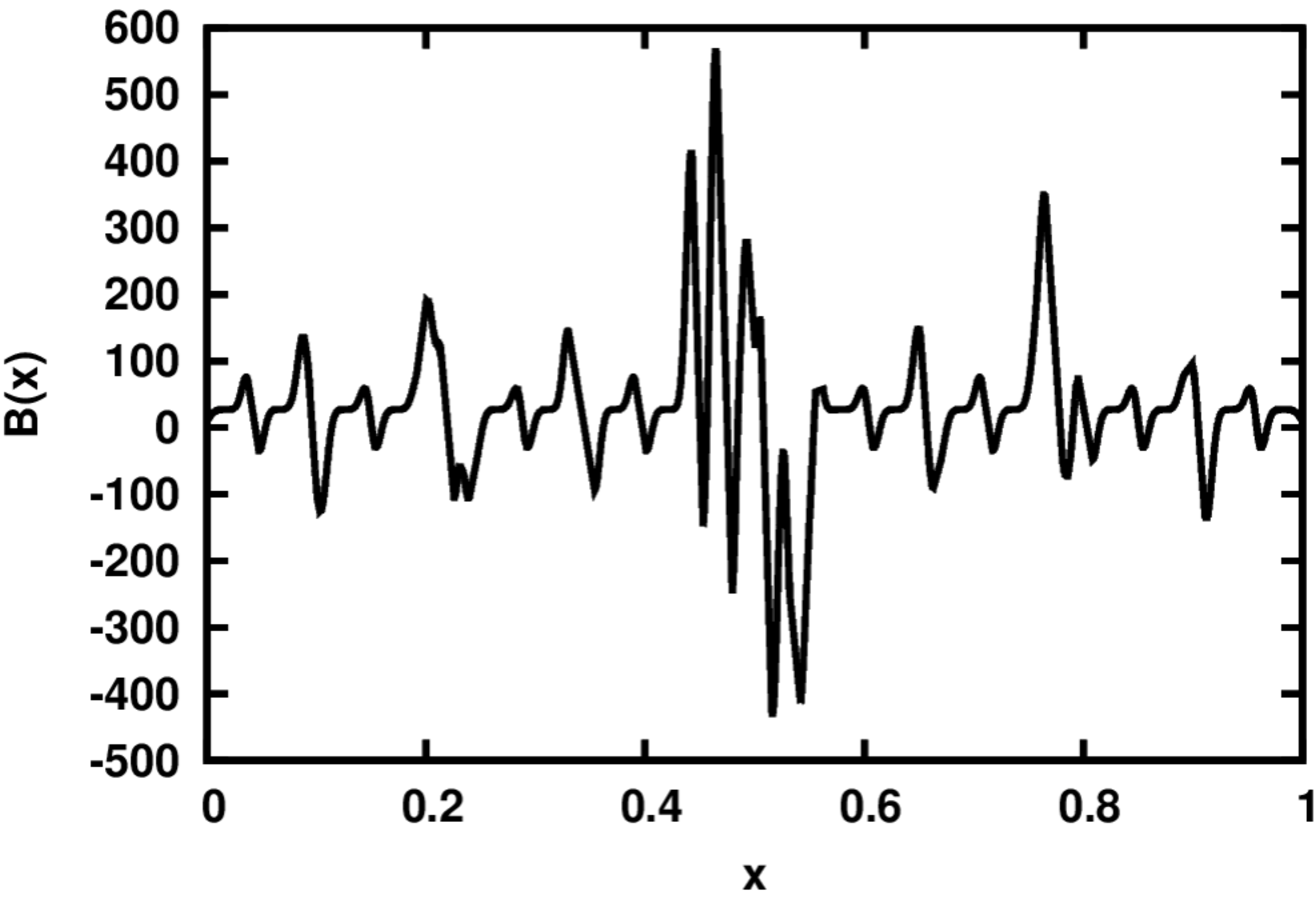}
 \caption{Eigenfunctions for the 4-strip map for $\Rm = 100,1000,10^4,10^5$
(top left to bottom right panels). We have adopted $\alpha = \delta = 7/16$ and 
$\beta = \gamma = 1/16$.}
\label{4strip}
 \end{figure}

This is clear from Fig.~\ref{2strip}: the eigenfunction of the magnetic field 
develops a finer and finer structures with increasing $\Rm$
(from $10^2$ to $10^5$). 
The eigenfunctions shown in Fig.~\ref{2strip} 
for different values of $\Rm$ match those in Fig.6 of \citet{FO90}
for the same $\alpha=0.4$.
The eigenfunctions for the four-strip map with $\Rm$ 
ranging from $10^2$ to $10^5$ 
are shown in Fig.~\ref{4strip}. 
Besides showing features similar to the two-strip maps, 
the eigenfunctions of the four-strip map have fine-scale 
reversals of magnetic field. 
These could be thought of as an analogue of the field reversals
seen in the DNS of fluctuation dynamos \citep{Schek04,BS05}.

\begin{figure}
\begin{center}
 \includegraphics[width=13cm,height=7cm]{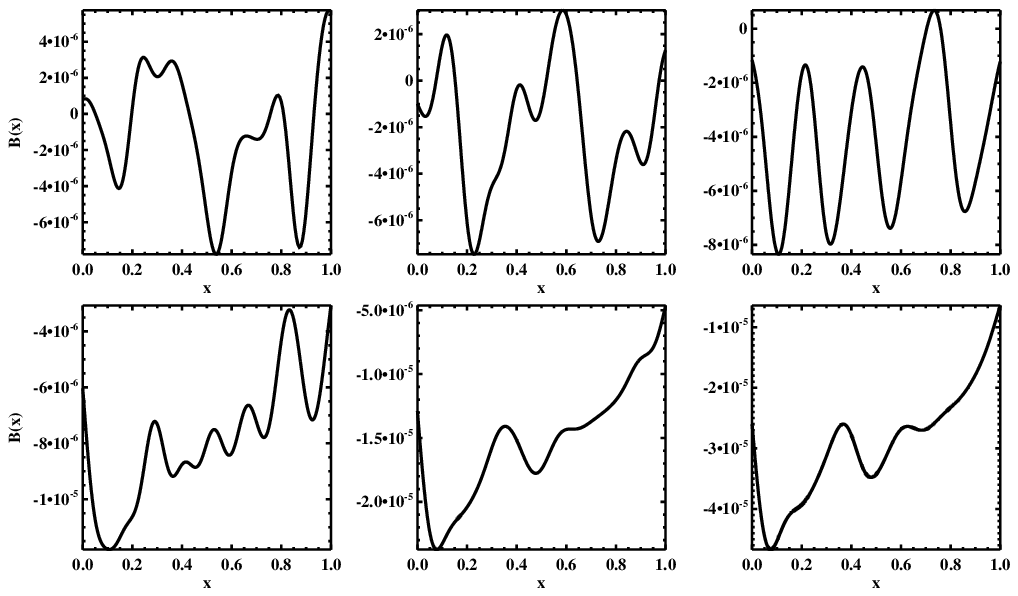}
 \caption{$B(x)$ evolution starting from a random initial seed field, for $\Rm=1000$.
 The iteration number increases from top left to bottom right. The $B(x)$ 
 in the bottom rightmost panel is the same eigenfunction 
  as $Rm=1000$ case is Fig.~\ref{2strip}, but with a negative amplitude}
 \label{randomSeed}
\end{center}
\end{figure}
An interesting case is when we use a random initial seed field, instead
of a uniform one. In the Fig.~\ref{randomSeed}, we show the evolution of
such a run, starting from top left to bottom right. It can be seen that as the
run progresses, the initial random field with both positive and negative values
changes to an eigenfunction which is entirely negatively valued. This is clear
on noting that the zero on the ordinate axis rapidly moves up showing that $B(x)$
becomes more and more negative as it latches on to the eigenfunction. 
The $B(x)$ in the bottom rightmost panel in Fig.~\ref{randomSeed} is the same eigenfunction 
as $Rm=1000$ case is Fig.~\ref{2strip}, but with a negative amplitude.
It could as well have become entirely positively valued depending on the initial seed field.
This example also illustrates the fact that the eigenfunction of the map is 
realised independent of the initial conditions.

\begin{figure}
\begin{center}
 \includegraphics[width=10cm,height=6cm]{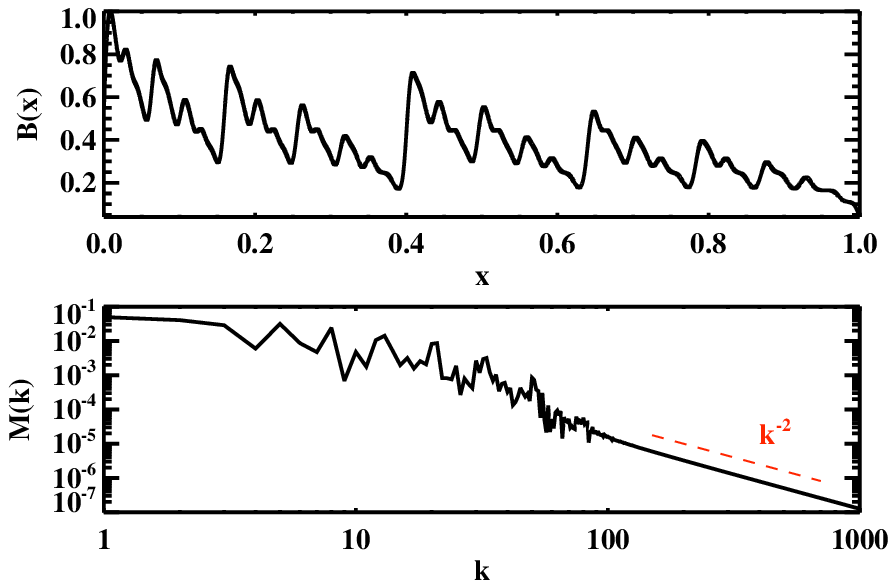}
 \caption{In the top panel, we show the eigenfunction $B(x)$ for a 2-strip map dynamo 
 for $\Rm=10^6$. In the bottom panel, we show the corresponding power spectrum $M(k)$.}
 \label{fteg}
\end{center}
\end{figure}

\subsection{Fourier Analysis of the magnetic field}

We can also calculate the Fourier series for $B(x)$ to study the distribution of the magnetic
power on different scales. The Fourier series is given by,
\begin{equation}
\tilde{B}(k)=\int_0^1 B(x)~e^{-2\pi i x k}~dx
\label{fsb}
\end{equation}
where, $k$ is wavenumber conjugate to $x$.
Then the magnetic power spectrum can be defined as $M(k)=\vert\tilde{B}(k)\tilde{B}^*(k)\vert/2$.

In the Fig.~\ref{fteg}, we show $M(k)$ corresponding to the eigenfunction B(x)
resulting from the  two-strip map dynamo with $\Rm=10^5$.
The log-log plot does not show $M(k)$ for $k=0$ which holds a significant
amount of total power. However in the given context, the $k=0$ component is just an 
overall constant factor in the eigenfunction and we would instead like 
to focus on the distribution of the power on smaller scales.
We find that $M(k)$ is non-smooth and can vary sharply within adjacent values of $k$.
This is due to the fractal nature of the eigenfunction, $B(x)$.
For fractal functions, several intermediate co-efficients in their Fourier series
are expected to be zero \citep{Korner88}.
 
\begin{figure}
 \includegraphics[width=6.6cm,height=10cm]{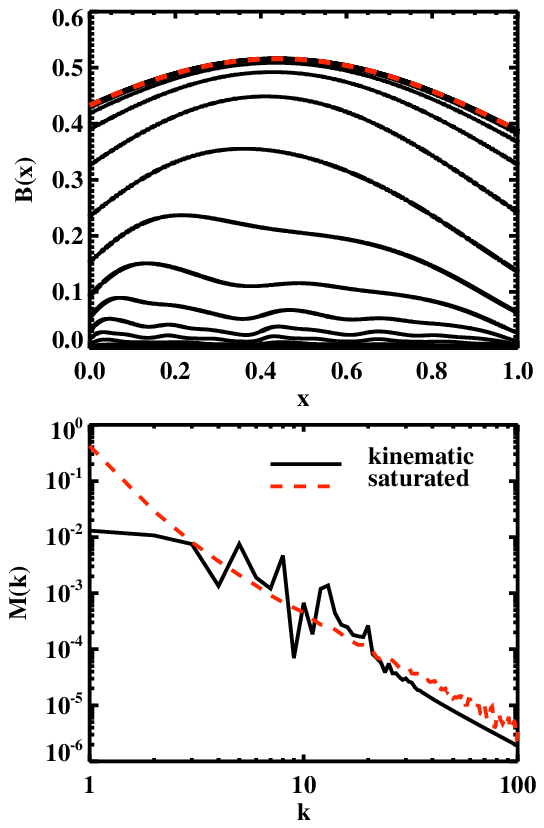}
 \includegraphics[width=6.5cm,height=10cm]{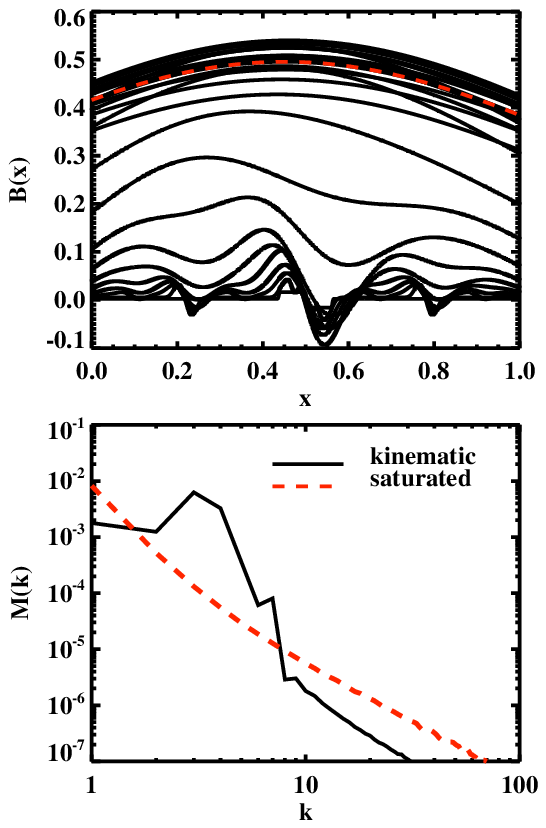}
 \caption{The nonlinear evolution of $B(x)$, 
due to saturation by increasing resistivity.
The top left panel shows 2-strip map evolution adopting $\alpha = 0.4$ and initial $\Rm = 10^4$.
The top right panel shows 4-strip map evolution with $\alpha = \delta = 7/16$, $\beta = \gamma = 1/16$ 
and initial $\Rm = 10^6$. We see that the eigenfunction in both cases  
is driven to that of the marginal eigenfunction corresponding
to $\Rm=\Rmc$ on saturation.
The bottom two panels show the magnetic power spectrum, $M(k)$, 
for the two cases respectively. The bold black line shows $M(k)$
in kinematic regime and the red dashed line is the saturated case.}

\label{decrm}
\end{figure}

Note that $M(k)$ at large k, follows as $1/k^2$. This is because at large k, 
the sines and cosines in Eq.~\ref{fsb} would sample a nearly constant part of $B(x)$
(or a ramp function in this case). And the Fourier transform of the ramp function is $\propto 1/k^2$.
In figures~\ref{decrm},\ref{decstr} and \ref{c1c2}, we show $M(k)$ for both two-strip
and four-strip cases for different $\Rm$ in kinematic stage and compare it
to the saturated one. We will say more about these figures below in the following section.

\section{Saturation of STF map dynamos}

Saturation of dynamos can occur in several different ways. 
Possibilities include the re-normalization and increase 
of the effective resistivity due to Lorentz forces \citep{S99} or 
the decreased stretching efficiency \citep{Schek04}. We model 
these in simple ways below to study the saturation of the map dynamos. 
We will now set the absolute value of the 
saturated magnetic field strength to be of order unity,
and therefore start with an initial seed field of $10^{-4}$.

\subsection{Saturation by decreasing $\Rm$}

As one possibility, consider saturating the dynamo by 
nonlinear increase of the effective resistivity as in the 
ambipolar drift model of \citet{S99}. In this model as the magnetic field
grows and Lorentz forces become important, the effective resistivity
becomes $\eta = \eta_0 + \tau B_{rms}^2/4\pi\rho$.
Here $\eta_0$ is the microscopic resistivity, $\tau$ is a response time, 
$\rho$ is the fluid density and $B_{rms}$ the rms value of $\vert{\bf B}\vert$. 
Multiplying and dividing the second term
in $\eta$ by $v^2$, where $v$ is the rms turbulent velocity, we
can rewrite this as $\eta = \eta_0( 1+ R_{M0} B_{rms}^2/B_{eq}^2)$,
where $R_{M0} = v^2\tau/\eta_0$ and $B_{eq}^2 = 4\pi\rho v^2$.
Thus $\eta_0/\eta = \Rm/R_{M0} = ( 1+ R_{M0} B_{rms}^2/B_{eq}^2)^{-1}$.
We adopt such a picture for the map dynamo as well
and model the nonlinear effect of the growing field by
varying $\Rm$ at every iteration as,
\begin{equation}
 \Rm = \frac{R_{M0}}{1 + R_{M0} B_{rms}^2}.
\label{Rmnew}
\end{equation}
Again $R_{M0}$ is the initial value of $\Rm$ for the map 
and $B_{rms}^2$ is now the average
value of $B^2(x)$ at any iteration (or time),
taken to be normalised to the equipartition value.  

This form models the possible increase of the renormalized 
resistivity due to Lorentz forces.
\begin{figure}
\begin{center}
 \includegraphics[width=12cm,height=8.5cm]{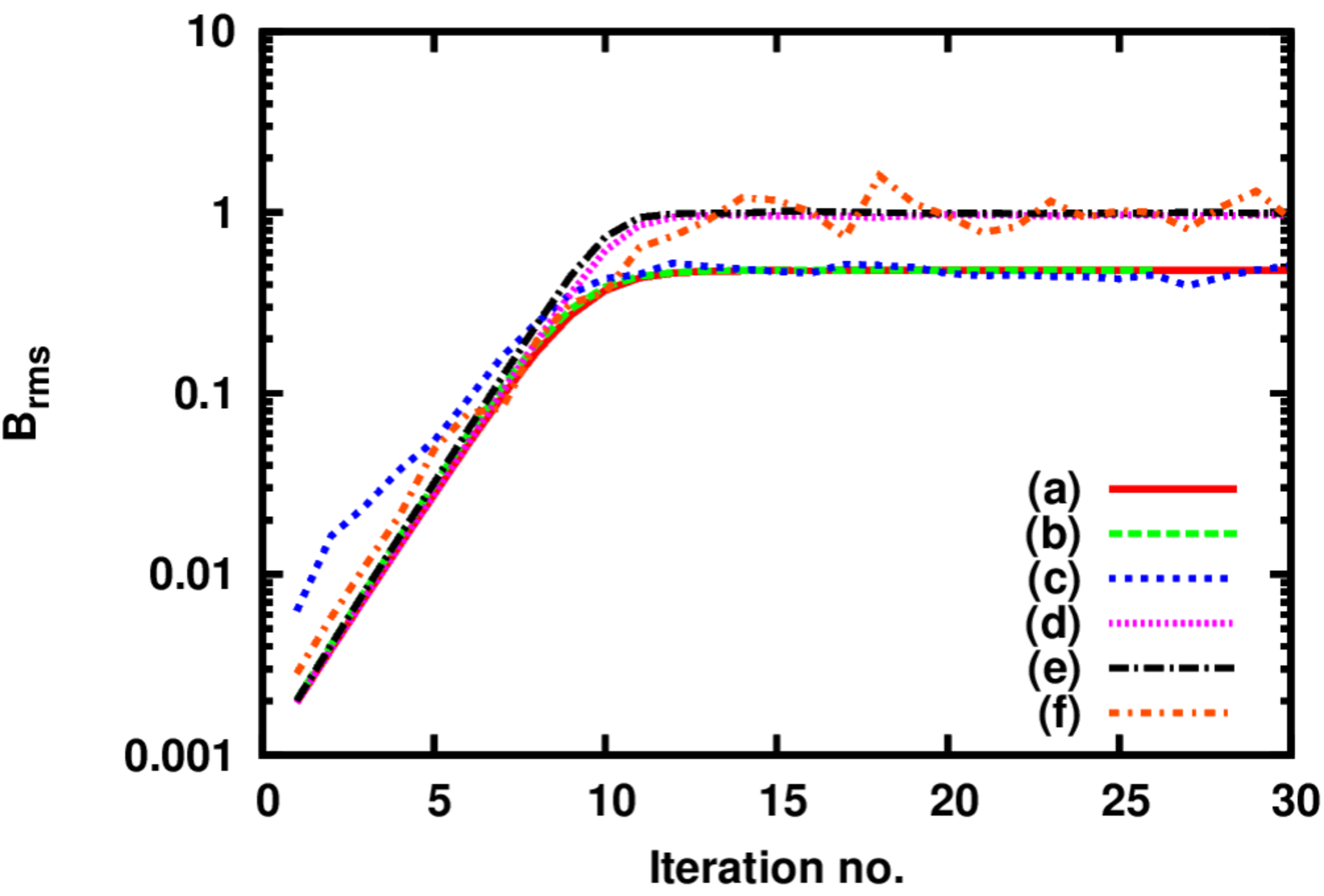}
 \caption{Comparison of $B_{rms}$ for various cases, (a): 2 strip map with $\Rm = 1000$, saturation by decreasing $\Rm$,
 (b): 2 strip map with $\Rm = 10^4$, saturation by decreasing $\Rm$,(c): 4 strip map with $\Rm = 10^6$, saturation by decreasing $\Rm$,
 (d): 2 strip map with $\Rm = 1000$, saturation by decreasing stretching, (e): 2 strip map with $\Rm = 10^4$, saturation by decreasing stretching,
 (f): 4 strip map with $\Rm = 1000$, saturation by decreasing stretching}
\label{brmsevol1}
\end{center}
 \end{figure}

The time evolution of $B(x)$ is shown in 
the left and right panels of Fig.~\ref{decrm} 
for 2-strip, $R_{M0} = 10^4$ and  4-strip, $R_{M0} = 10^6$ respectively.
The evolution of corresponding $B_{rms}$ is shown in Fig.~\ref{brmsevol1} 
as case (a) and case (c) respectively. We see from Fig.~\ref{brmsevol1} 
that $B_{rms}$ indeed saturates after about $7-10$ iterations, to a 
value of order unity. Moreover, on comparing left panel of Fig.~\ref{decrm}
with left panel of Fig.~\ref{rmcrit} and right panel of Fig.~\ref{decrm}
with right panel of Fig.~\ref{rmcrit}, it is clear that the function $B(x)$ representing 
the saturated state, is of the same form as $B(x)$ 
for the critical $\Rm$($\sim 4.35$).
Further in the saturated stage the $\Rm$ given 
by the Eq.~\ref{Rmnew}, also settles to $\Rsat=\Rmc\sim4$,
using $B_{rms}\sim0.5$ and $R_{M0}=1000$. 

In Fig.~\ref{decrm}, we also show the corresponding magnetic power
spectrum, $M(k)$, where the solid black curve is from the kinematic stage
and the red dashed line is for the saturated field.
It can be seen that the kinematic field exhibits peaks on smaller scales,
around $k\sim 5-20$ in 2-strip case and $k\sim 3-10$ in the 4-strip case.
The peaks seen on smaller scales (or larger k) in kinematic stage seem to be smoothed
out by saturation with an increase in the power in $k\sim 1-2$.
Note that in both stages, however, the maximum power is in $k=0$
constant component, which does not appear in such a log-log plot. 
Thus in the case where saturation is obtained by renormalization 
of $\Rm$, we find that the saturated state has the same spatial
structure as the marginal state of the kinematic map dynamo. 
Such a result is similar to the saturation behaviour obtained 
in \cite{S99} and perhaps in the simulations of \cite{HBD04} for 
the $\Pm = 1$ fluctuation dynamo.

\begin{figure}
 \includegraphics[width=6.5cm,height=10cm]{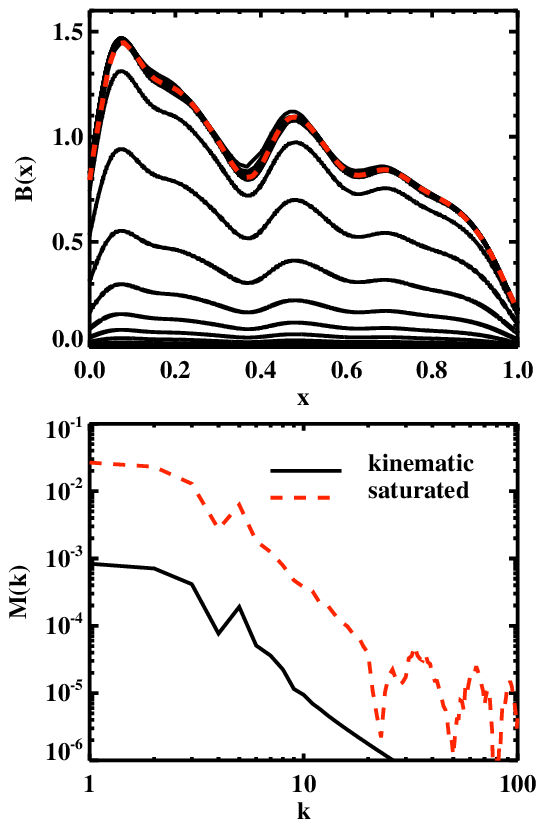}
 \includegraphics[width=6.5cm,height=10cm]{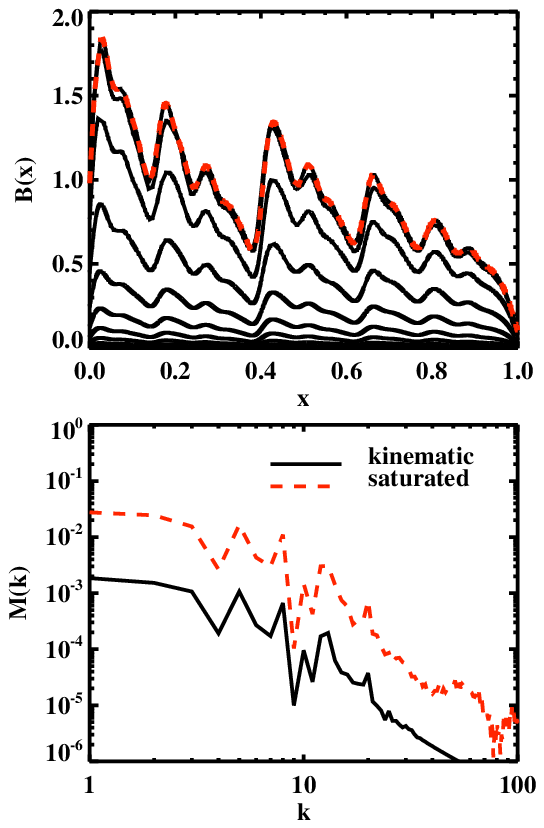}
 \caption{The nonlinear evolution of $B(x)$ 
of the 2-strip map, due to saturation by reduced stretching,
adopting initial $\alpha_0 = 0.4$ and $\Rm = 1000$ (left top panel)
and $\Rm=10^4$ (right top panel). The red dashed line show $B(x)$ in saturated state.
The bottom two panels show the magnetic power spectrum, $M(k)$, 
for the two cases respectively. The bold black line shows $M(k)$
in kinematic regime and the red dashed line is the saturated case.}
\label{decstr}
\end{figure}

\begin{figure}
 \includegraphics[width=12cm,height=8.25cm]{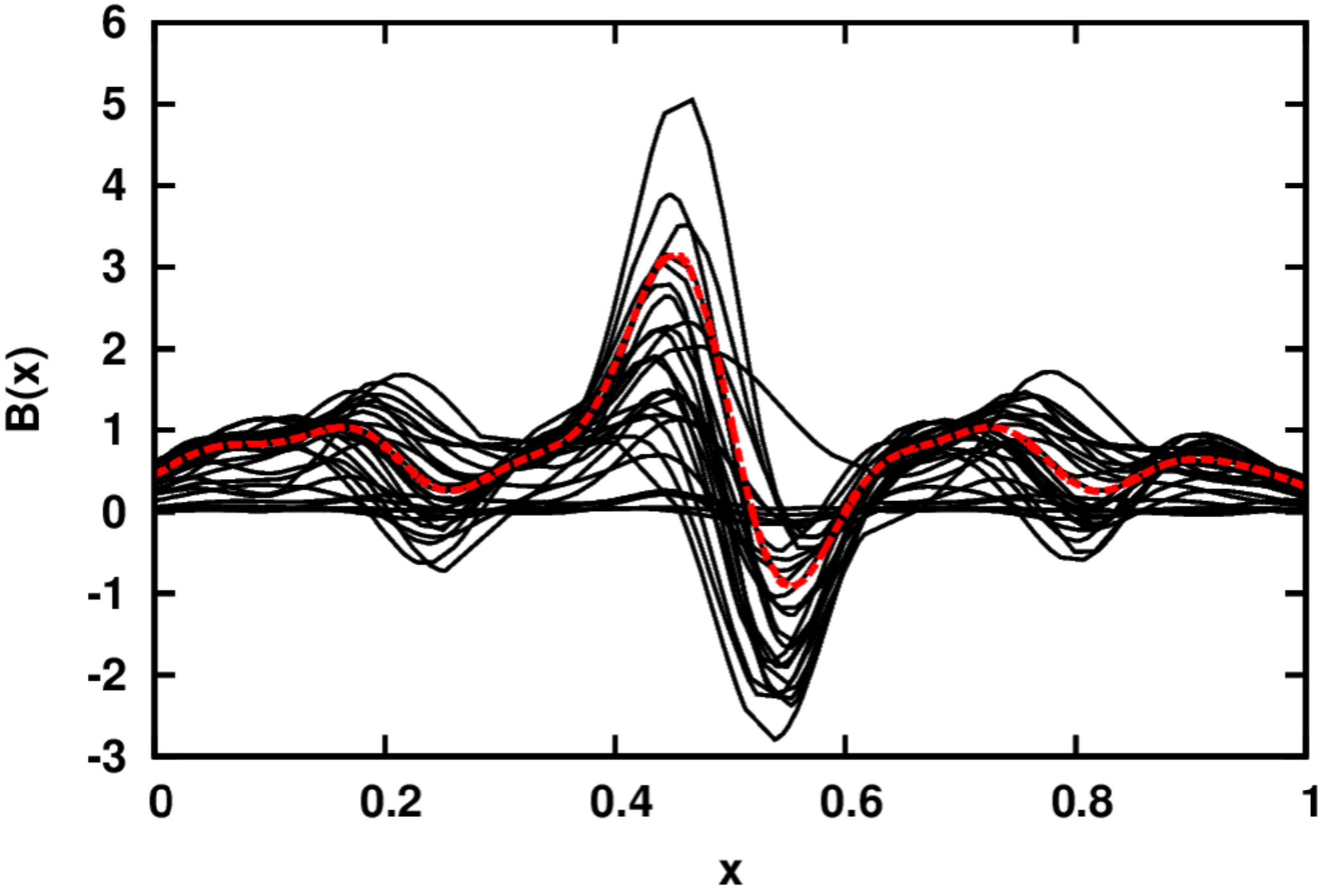}
 \caption{Same as in Fig.~\ref{decstr} but now for the 4-strip map
adopting an initial $\alpha_0 = \delta_0 = 7/16, \beta_0 = \gamma_0 = 1/16$ 
and $\Rm = 1000$. The eigenfunction on saturation oscillates with a structure similar to the kinematic eigenfunction. The shape 
of kinematic eigen function is shown in dashed (red) line.}
\label{4decstr}
 \end{figure}

\subsection{Saturation by decreasing stretching}

\subsubsection{Without changing the map}

As the magnetic field strength increases, the effect of the Lorentz force
would be to make it more difficult to amplify the field by
stretching. Therefore, another way to achieve saturation in the
maps would be to
decrease the field amplification factor 
$\propto 1/\alpha$ as a function of $B_{rms}$. 
We model this effect by multiplying $\alpha$, $\beta$ in 
Eq.~\ref{eq.3} for the 2-strip map and 
$\alpha$, $\beta$, $\gamma$, $\delta$ in 
Eq.~\ref{eq.6} for the 4-strip map by $(1 +B_{rms}^2)$. For example we adopt 
\begin{equation}
 \alpha = \alpha_0 (1 + B_{rms}^2),
\label{stretch}
\end{equation}
where $\alpha_0$ is the initial value of $\alpha$.
Note that to begin with we still leave the mapping of $(x_n,y_n) \to (x_{n+1},y_{n+1})$
described by Eqs.~\ref{eq.1}, \ref{eq.2}, \ref{eq.4} and \ref{eq.5}
as before, described by the initial $\alpha_0$, $\beta_0$, $\gamma_0$ and 
$\delta_0$. Thus the unit square in the $x-y$ plane is still mapped
to the unit square, but the amplification by stretching becomes
progressively inefficient as $B_{rms}$ grows.

Note that as one reduces stretching by a factor 
$f_0=1/(1 + B_{rms}^2)$ and applies the STF map to a flux tube, 
its final radius $R$ would decrease by $f_0$ while its
cross sectional area $A$ would increase by $1/f_0$.
In principle the physical dimensions of the unit square that we use 
to represent the flux tube would then change accordingly and it will 
become a rectangle with its y-dimension (analogous to the length of the flux tube) 
reduced by $f_0$ and $x$ dimension (analogous to the cross sectional 
area of the flux tube) increased by $1/f_0$. 
However, we are thinking of the unit square used in the map
as representing the normalised flux tube radius and the normalised
cross-sectional area. It is in this sense that the unit square
is mapped onto itself, even though the degree of stretching is reducing
with the growth of the field.
We also now do not change the value of $\Rm$ during the diffusive step.

The result of the such reduced stretching for the 2-strip map is shown in 
left and right top panels of 
Fig.~\ref{decstr} 
for $\Rm = 1000$ and $\Rm = 10^4$ respectively. 
The corresponding $B_{rms}$, whose evolution is shown 
in Fig.~\ref{brmsevol1} as cases (d) and (e),
saturates after about 10 iterations.
It is clear from comparing Fig.~\ref{decstr} 
with Fig.~\ref{2strip}, that the distribution $B(x)$ representing the 
saturated state in this case retains the complex structure of the
growing kinematic eigenfunction and is quite different from the
case of saturation by increased resistivity. 
It is thus not the marginal eigenfunction. 
In Fig.~\ref{decstr}, we also show the magnetic power spectrum $M(k)$
in the bottom two panels. Here again we see peaks on small scales, at $k\sim6$
for $\Rm=1000$ and $k\sim 4-20$ for $\Rm=10^4$. The saturated spectrum here 
matches with the one in the kinematic stage.
This saturation behaviour is similar to what 
is argued by \cite{Schek04} for the fluctuation dynamo with large $\Pm$,
that the field in the saturated state appears to be qualitatively similar to
the growing field in the kinematic stage.

We have also employed a similar scheme for the 4-strip map with $\Rm = 1000$. 
The results shown in Fig.~\ref{4decstr} indicate a very different saturation behaviour. Now, the saturated eigenfunction $B(x)$ oscillates
with time although its form is similar to kinematic eigenfunction.
Also $B_{rms}$ oscillates about a steady value around unity, as
can be seen in Fig.~\ref{brmsevol1} (case (f)). This seems to indicate that 
if reversals are present, the saturated eigenfunction may never settle 
to a unique form.

\subsubsection{Changing the map}

\begin{figure}
\begin{center}
 \includegraphics[width=12cm,height=8.5cm]{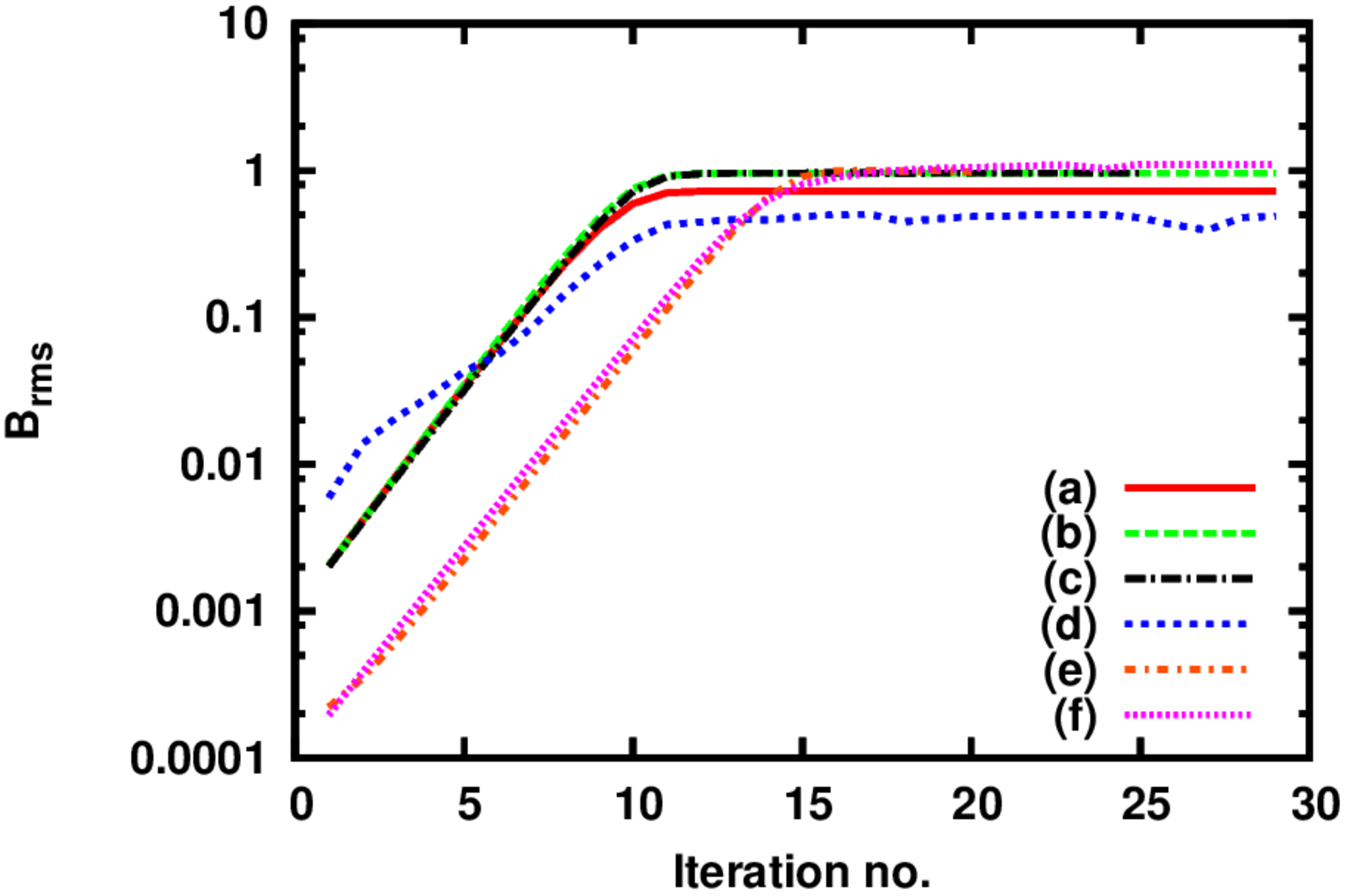}
 \caption{Comparison of $B_{rms}$ for various cases,
 (a), (b), (c) and (d) show curves for saturation by combining decreasing both $\Rm$ and stretching.
 (a): 2 strip map with initial $\Rm=10^6$, $c1=0.1$, $c2=0.9$, 
 (b): 2 strip map with initial $\Rm=10^6$, $c1=0.001$, $c2=0.999$,
 (c): 2 strip map with initial $\Rm=10^4$, $c1=0.001$, $c2=0.999$,
 (d): 4 strip map with initial $\Rm=10^5$, $c1=0.001$, $c2=0.999$, 
 (e): 2 strip map with initial $\Rm=10^3$, saturation by decreasing stretching (by changing map),
 (f): 2 strip map with initial $\Rm=10^3$, saturation by decreasing stretching (by changing map)
      and also decreasing $\Rm$, adopting $c_1=0.1$ and $c_2=0.9$.}
\label{brmsevol2}
\end{center}
 \end{figure}

\begin{figure}
 \includegraphics[width=6.5cm,height=5cm]{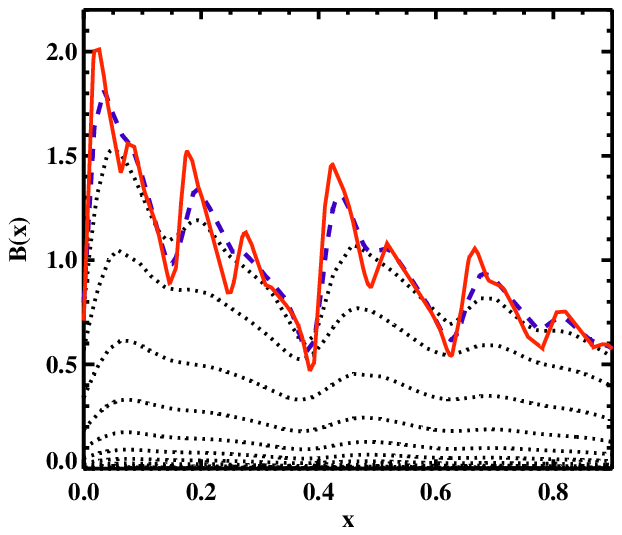}
 \includegraphics[width=6.5cm,height=5cm]{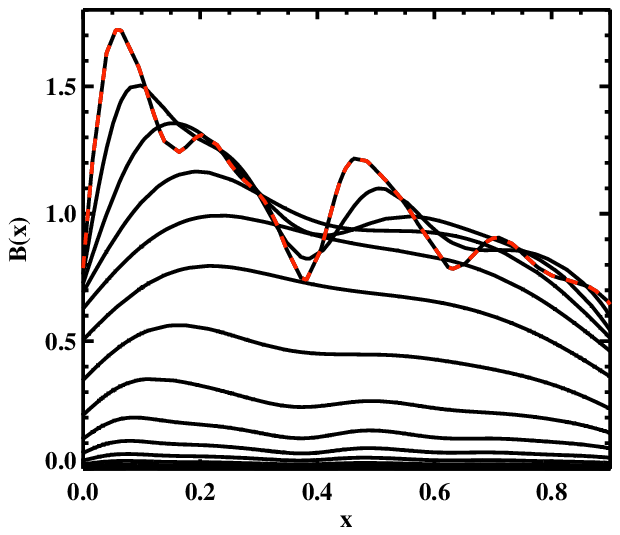}
 \caption{The evolution of $B(x)$ in a 2-strip map dynamo, $\Rm=1000$ in 2 cases.
 The left panel shows saturation by only reduced stretching with changed map.
 The last two iteration have been shown as dashed blue line and solid red line.
 And the right panel shows saturation by reduced stretching with changed map 
 along with decreasing $\Rm$, with $c_1=0.1$ and $c_2=0.9$. The final iteration is shown
 in dashed red line.}
\label{mapchange}
 \end{figure}

On reduced stretching, as explained earlier, the physical
dimensions of unit square map will change to become a rectangle.
In the previous subsection, such a changed map is renormalised to unit square before the 
diffusion is carried out, thus effectively not changing the map.
We now explore the consequences of renormalising the map to unit square only after the diffusion step;
before we carry out further STF mapping. This renormalization process is
to ensure that the next STF mapping can be done, as before, on a unit square.
Such a modified mapping for the 2-strip case can be given as,
\begin{equation}
  x_{n+1} = \left\{
     \begin{array}{ll}
      \alpha_1 x_n &: y_n < \alpha; \\ 
      \beta_1 x_n + \alpha_1 &:  y_n > \alpha; \\
     \end{array} \\
   \right. 
\label{eq.1} 
\end{equation}
\begin{equation}
 y_{n+1} = \left\{
     \begin{array}{ll}
       y_n/\alpha_1 &: y_n < \alpha; \\ 
       (y_n - \alpha)/\beta_1 &:  y_n > \alpha; \\
     \end{array}
   \right. 
\label{eq.2} 
\end{equation}
The corresponding amplification of magnetic field in the region is given as: 
\begin{equation} 
B_{n+1}(x_{n+1}) = \left\{
     \begin{array}{ll}
      B_n(x_n)/\alpha_2 &: x_n < \alpha_1; \\ 
      B_n(x_n)/\beta_2 &:  x_n > \alpha_1; \\
     \end{array}
   \right. 
\label{eq.3} 
\end{equation}
where initially, both $\alpha_1$ and $\alpha_2$ are $\alpha_0$.
After the diffusion step, once we evaluate the current $B_{rms}$,
we estimate $\alpha_1= \alpha_1 (1 + B_{rms}^2)$
and $ \alpha_2=\alpha_0 (1 + B_{rms}^2)$. 

This implies that we are effectively
working with the rectangle, which gets elongated further and further as the dynamo progresses.
In Fig.~\ref{mapchange}, the left panel shows the evolution of a 2-strip, $\Rm=1000$ run with such a
saturation mechanism. We can see that the last few iterations overlap indicating the onset of
saturation. 
While the magnetic energy saturates as seen in Fig.~\ref{brmsevol2} in curve (e),
such a process eventually becomes unstable in $B(x)$. 
This is because as the run progresses, the map grows in $x$-direction 
and diminishes continuously in $y$-direction, thus making the diffusion 
step progressively inefficient. This tends to a scenario wherein the 
effective $\Rm$ for the map keeps increasing leading to the ideal case in Fig~\ref{fig.1}.
The $B(x)$ seems to acquire more and more structures,
as can be seen in the final two curves in dashed blue and solid red in Fig.~\ref{mapchange} (left panel), 
reflecting the scenario of growing effective $\Rm$. 

In the flux tube picture, the flux tube will keep thickening and simultaneously grow 
smaller in size (radius). But at some point, this process will have to stop 
when both dimensions become comparable. Of course, the other possibility is that
when the field grows to sufficiently high values, the tension in the flux tube 
will not allow further twisting and folding. 
One way of implementing a suppression 
in twisting and folding is to stop the map when the $B_{rms}$ exceeds a threshold and allow for only diffusion.
Although we have not explicitly shown this here, we expect that the $B(x)$ will freeze
to its form when it first crosses the threshold in $B_{rms}$ thus retaining some structures
from the kinematic eigenfunction.

\subsection{Saturation by combining decreasing both $\Rm$ and stretching}

We now consider the saturation of the STF map dynamo, when the
effects of decreasing effective $\Rm$ (due to increasing renormalised
resistivity) and  decreasing stretching efficiency are combined. 
We model this by introducing efficiency parameters $c_1$ and $c_2$
into Eqs.~\ref{Rmnew} and \ref{stretch}. We adopt 
\begin{equation}
\Rm = \frac{R_{M0}}{1 + c_1 R_{M0} B_{rms}^2},
\quad 
\alpha = \alpha_0 (1 + c_2 B_{rms}^2).
\label{combine}
\end{equation}
For $c_1=1$ and $c_2=0$ we have saturation purely by the nonlinear 
decrease of $\Rm$, while $c_1=0$ and $c_2=1$ corresponds to 
the case when saturation occurs purely due to reduced stretching.
We consider now the intermediate case where both $c_1$ and $c_2$ are 
non zero.

\begin{figure}
 \includegraphics[width=6.5cm,height=10cm]{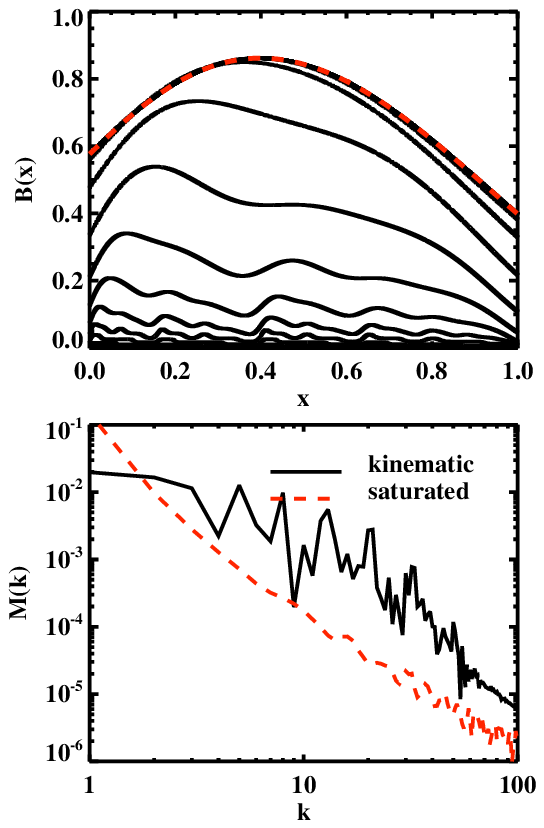}
 \includegraphics[width=6.5cm,height=10cm]{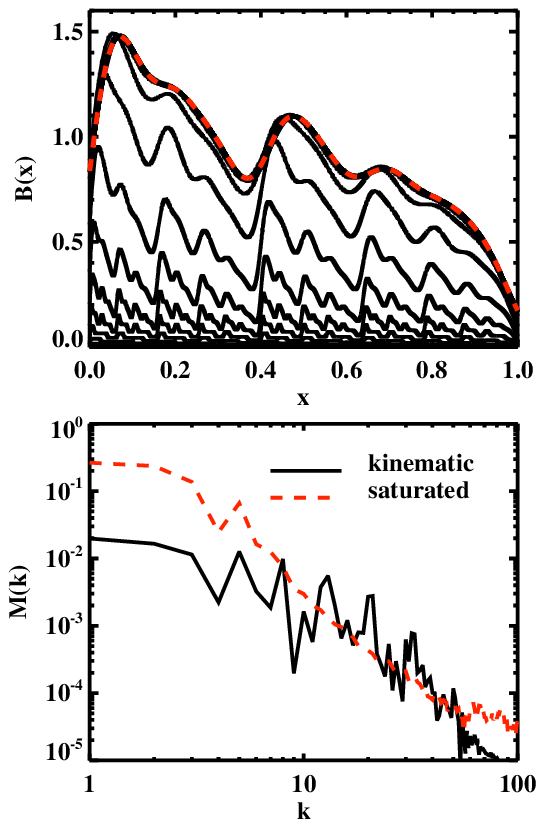}
 \caption{The nonlinear evolution of $B(x)$ 
of the 2-strip map, due to saturation by combining both increased
diffusion and reduced stretching (without changing the map),
adopting initial $\alpha_0 = 0.4$ and $\Rm = 10^6$. The top left panel
shows the result of adopting $c_1=0.1$ and $c_2=0.9$ (case A), while the top right
panel shows the case when $c_1=0.001$ and $c_2=0.999$ (case B). The effective
magnetic Reynolds number at saturation is $\Rsat=18$ and $\Rsat=1066$, 
for case A and B respectively. The corresponding kinematic eigenfunctions
are shown as dashed lines in the figure. 
We see that the the shape of $B(x)$ at saturation closely matches that
of the kinematic eigenfunction for $\Rm=\Rsat$.
The bottom two panels show the magnetic power spectrum, $M(k)$, 
for the two cases respectively. The bold black line shows $M(k)$
in kinematic regime and the red dashed line is the saturated case.}
 \label{c1c2}
\end{figure}

The evolution of $B(x,t)$ of the 2-strip map, adopting initial 
$\alpha_0 = 0.4$ and $R_{M0} = 10^6$ is shown in Fig.~\ref{c1c2}, 
for two cases. The top left panel
shows the result of adopting $c_1=0.1$ and $c_2=0.9$ (case A), while the top right
shows the case when $c_1=0.001$ and $c_2=0.999$ (case B). The effective
magnetic Reynolds numbers at saturation have now reduced from 
$R_{M0} = 10^6$ to $\Rm=\Rsat=18$ and $\Rm=\Rsat=1066$, 
for cases A and B respectively. The corresponding kinematic eigenfunctions 
are shown as dashed lines in the figure. 
Remarkably, we see from Fig.~\ref{c1c2} that the the shape of $B(x)$ at 
saturation for both cases, now closely matches that of the corresponding 
kinematic eigenfunction for $\Rm=\Rsat$. Thus when both the effective 
diffusion of the field and the stretching efficiency are affected by 
Lorentz forces, as would perhaps be most realistic, then the dynamo 
saturates with an intermediate spatial structure; that of the
kinematic eigenfunction with $\Rm=\Rsat$ with $R_{M0} > \Rsat > \Rmc$. 
$\Rsat$ is determined by the 
relative importance of the increased diffusion versus the reduced stretching.
However a change in $c_1$ has a more dramatic effect than an equal
change in $c_2$, as $c_1$ appears in an exponential function,
the Gaussian in Eq.~\ref{convolve} involved in convolution to 
incorporate resistive effects.  
In Fig.~\ref{brmsevol2}, we show saturation levels of the two runs in curves
(a) and (b), which are close to a value of order unity. We also show the curve (c) for a run with 
lower $\Rm=10^4$ which saturates to a similar level.

We also show the magnetic power spectrum, $M(k)$, in the bottom two 
panels in Fig.~\ref{c1c2}. For the case A, the saturated spectrum is smooth
and increases monotonically towards smaller $k$.
In case B, the saturated spectrum seems to still retain some peaks near $k\sim5$
and is flatter near $k\sim1-3$, unlike case A. Thus reflecting the nature of field
as expected for their corresponding lower and higher $\Rsat$ of 18 and 1066, respectively.

In the 4-strip map dynamo, this kind of saturation mechanism,
with co-efficients, $c_1=0.001$ and $c_2=0.999$, leads to large oscillations 
in the form of $B(x)$ similar to the case when the saturation was only by reduced stretching.
The field gets quite ordered by saturation as a result of inclusion of 
the mechanism increased resistivity even though the value of $c_1$ is very small.
These oscillations are also reflected in the evolution curve of $B_{rms}$
in the Fig.~\ref{brmsevol2} in curve (d).

Finally, we have also considered a case where we combine the two saturation mechanisms
and at the same time change the map. The resulting $B(x)$ evolution is shown in the right panel
of the Fig.~\ref{mapchange} adopting $c_1=0.1$ and $c_2=0.9$, with $\Rm=1000$, for a 2-strip map dynamo.
Compared to the case in the left panel in Fig.~\ref{mapchange}, $B(x)$ saturates to a smoother kinematic 
eigenfunction, but then, the diffusion becomes less and less important,
causing the field to develop finer scale structures, while its rms value
still maintains a steady state as seen in curve (f) of Fig.\ref{brmsevol2}.

\section{Conclusions}

We have explored here the evolution and saturation behaviour of map dynamos 
used by \citet{FO88,FO90} to model Zeldovich's STF dynamo. One of our aims is 
to use these simpler systems to develop some intuitive
understanding of the saturation behaviour of more realistic 
fluctuation dynamos. The use of maps also allows one to analyze dynamos 
with very high values of $\Rm$, much higher than what can be achieved in DNS.

We have considered in particular two types of the generalized Baker's maps,
the 2-strip map where there is constant constructive folding
and the 4-strip map which allows the possibility of reversal of the
field.  In the absence of diffusion, the magnetic field $B(x)$ generated
by the map dynamo develop fractal structures \citep{FO88,FO90}. 
On including a diffusive step in the map, parameterised by the magnetic
Reynolds number $\Rm$, we find that the magnetic field $B(x)$ 
latches on to an eigenfunction of the map dynamo and is amplified,
only above a critical $\Rm=\Rmc \sim 4$ for both types of dynamos. 
The spatial structure of the growing $B(x)$  also becomes shape invariant
(with iteration number), 
but whose complexity 
increases with increasing $\Rm$. 
Such an eigenfunction is obtained independent of the initial field
configuration, as we illustrated with an initial random seed field.
The kinematic eigenfunction of the
4-strip map shows reversals of the field, whose number increases
with $\Rm$. These results are similar to those presented in \citet{FO88,FO90}
on the kinematic stage of the map dynamos, wherever the comparisons 
can be made. We have also illustrated both the kinematic and saturated structure of 
$B(x)$ by considering its power spectrum $M(k)$. In the kinematic case,
the fractal nature of $B(x)$ till the diffusive scale is reflected in sharp peaks in $k$-space.

We then explored 
different ways by which the STF map dynamos 
could saturate. 
Saturation can occur due to a renormalized increase of the effective 
resistivity or by decreasing $\Rm$. Such an effect obtains in a model
where the Lorentz forces leads to a `ambipolar'-type nonlinear drift 
velocity \citep{S99}. 
For both the 2-strip map and the 4-strip map which includes field reversals,
$B(x)$ in the saturated state goes back to the 
marginal eigenfunction, which would obtain for the critical $\Rm=\Rmc$.
The structure developed during the kinematic stage is lost on
saturation and thus one can conclude that the
energy is being effectively transferred to the larger scales
due to non-linear evolution 
as can also be seen from their power spectra. This is analogous to the analytical results of 
\cite{S99} and the DNS results of \citet{HBD04,Eyink13,BS13}
for fluctuation dynamos with $\Pm=1$.

We have also explored the saturation of the dynamo, when the
effect of Lorentz forces is to  decrease the efficiency
of field amplification by stretching.
This is implemented in two ways, (i) by not changing the map
and (ii) when the map is changed to reflect the decreasing physical
length (radius) of the flux tube and its increasing thickness.

For the 2-strip map, in case (i), we show that $B(x)$ now saturates preserving
the structure of the kinematic eigenfunction 
got using the initial $\Rm=R_{M0}$.

In the case (ii), where we implement reduced stretching in the map,
the magnetic energy  (or $B_{rms}$) saturates, but
but the fractal structures in $B(x)$ keep growing as in the ideal limit.
Thus energy is still 
preserved at the smallest scales which 
survive
resistive diffusion. 
This is analogous to the results of \citet{Schek04} 
for the large $\Pm$ and small $\Rey$ fluctuation dynamo, where power on 
resistive scales appear to be preserved during saturation.
However an intermediate behaviour obtains when both saturation mechanisms
operate in tandem. The saturated $B(x)$ has now the spatial structure 
of kinematic eigenfunction with an intermediate $\Rm=\Rsat$ 
where $R_{M0} > \Rsat > \Rmc$. Even a small
increase in the effective diffusion with growing field, 
(with $c_1 \ll 1$) significantly smoothens the spatial structure of the field. 
For the 4-strip map, saturation due to decreased stretching
efficiency with or without increased diffusion leads to a more 
complicated behaviour, as now the saturated $B(x)$ oscillates 
with time, although with a structure similar to the kinematic eigenfunction.

One could have naively expected that $B(x)$ is driven to a
universal form on saturation. However, it appears that the 
field structure when dynamos saturate
is a nontrivial issue even for the simple map dynamos,
and depends on the exact manner of saturation.
The two natural possibilities, that the saturated eigenfunction
approaches the marginal eigenfunction or remains of the same
form as the kinematic eigenfunction,
are both realized for different modes of saturation.
If one takes a hint from such map dynamos,
then even for the fluctuation dynamo in a random flow,
the structure of the saturated state could depend on
the control parameters of the system, like $\Rm$, $\Rey$ and
$\Rmc$. It would be interesting to explore such issues further.
It would also be interesting to extend the map dynamos to incorporate
a range of length scales, so as to mimic more realistically a turbulent
flow with a range of eddy scales, and study their saturation behaviour.

\section*{Acknowledgments}
AS thanks IUCAA for hospitality during his visits there under
the Visiting Student Programme. PB acknowledges support from CSIR.
We acknowledge the use of the high performance computing facility
at IUCAA. We thank two referees for useful comments which has led to
many improvements in the paper and Prof.~A.~D.~Gangal for sharing his
thoughts on the Fourier analysis of fractals.
\bibliographystyle{jpp}

\appendix

\label{lastpage}
\end{document}